\documentclass[12pt]{article}
\usepackage{jheppub}
\usepackage{MnSymbol}

\def\Z{{\cal Z}}
\def\be{\begin{equation}}
\def\ee{\end{equation}}
\def\ba{\begin{aligned}}
\def\ea{\end{aligned}}
\def\ben{\begin{eqnarray}\displaystyle}
\def\een{\end{eqnarray}}

\def\nn{\nonumber}
\def\rd{{\rm d}}

\def\tm{{\tilde m}}
\def\te{{\tilde e}}

\preprint{
{\small{\textsf{QMUL-PH-11-07}}}}

\title{3D-partition functions on the sphere: exact evaluation and mirror symmetry}

\author[1]{Sergio Benvenuti}  \author[2]{Sara Pasquetti}

\affiliation[1]{Theoretical Physics Group, The Blackett Laboratory,\\
 Imperial College, Prince Consort Road, London SW7 2AZ, UK\\}

\affiliation[2]{School of Physics, Queen Mary University of London, \\
Mile End Road, London E1 4NS, UK\\}

\abstract{We study   $\mathcal{N}=4$ quiver theories on the three-sphere.
 We compute partition functions using  the localisation method by Kapustin et al. 
 solving exactly the matrix integrals at finite $N$, as functions of mass and Fayet-Iliopoulos parameters.
We find a simple explicit formula for the partition function of the quiver tail  $T(SU(N))$.
This formula  opens  the way for the analysis of star-shaped quivers and their mirrors
 (that are the Gaiotto-type theories arising from M5 branes on punctured Riemann surfaces).
We provide non-perturbative checks of mirror symmetry for infinite classes of theories and find the partition functions of  the $T_N$ theory, 
the building block of  generalised quiver theories.
}    

\begin{document}

\maketitle

\section{Introduction}

Over the last two years the new class of 4$d$ $\mathcal{N}=2$ super-conformal  gauge theories introduced by Gaiotto \cite{gaiotto}
has attracted much interest. 
 This class of theories can be viewed as arising from a super-conformal  $6d$  theory,
the world-volume theory of $N$ M5 branes, compactified 
on the product of  Minkowski $4d$  space and  a  genus $g$ Riemann surface with $n$ punctures. 
Each puncture is associated with a Young diagram specifying the behaviour of the fields at the puncture.
Any Riemann surface with punctures can be constructed by gluing  spheres with three punctures
through  tubes.  Correspondingly, a generalised quiver  theory can be constructed by taking a number of $T_N$ theories, 
the low-energy limit of N M5-branes on a sphere with three punctures,
and gauging together their flavour symmetries.
 The  $T_N$ theory is the building block of generalised quiver theories, 
it has $SU(N)^3$ flavour symmetry, no marginal couplings
 and does not admit a  Lagrangian description.
The Gaiotto construction  provides a new unifying framework for the study of S-duality of $\mathcal{N}=2$ theories
which encodes and  extends the early observation by Argyres and Seiberg \cite{as} who discovered the non-Lagrangian E6 theory \cite{mn}, in strong coupling limit of  the $SU(3)$ theory coupled to 6 fundamental hypers.

In this paper we study the 3d version of this class of generalised quiver  theories.
After $S_1$ compactification, 4d theories  flow to an IR point fixed leading to 
$\mathcal{N}=4$ super-conformal generalised quiver theories in 3d.
The vacuum moduli space of $\mathcal{N}=4$  theories in 3d
consists of  a  Coulomb and a Higgs branch
corresponding respectively to fluctuations of massless vector multiplets and hypermultiplets.
A very interesting duality acting on this moduli space is  mirror symmetry \cite{is} 
which exchanges the Higgs and Coulomb branches of mirror pairs of theories swapping 
mass parameters for the hypermultiplets   with Fayet-Iliopoulos (FI) parameters for the vector multiplets. 
 
In \cite{sici}, by generalising the construction of mirrors of standard gauge theories  involving 
 D3-branes suspended between 5-branes  given in \cite{hw}, it has been
 found  the mirror of generalised quiver theories.
 For a theory associated to a sphere with $k$ punctures the mirror theory is conjectured to be 
a star-shaped quiver with $k$ arms coupled to a central  $SU(N)$ node.
Interestingly  star-shaped theories, mirror of
generalised quiver theories including $T_N$ blocks,
 turn-out to be always weakly coupled and admit a Lagrangian description.
 In this paper we will compute  partition functions of this class of  3d $\mathcal{N}=4$ theories by means of localisation techniques.

The technique of localisation of supersymmetric partition functions involves the addition of a Q-exact operator to the action,
 which does not affect the path integral, but renders the 1-loop approximation exact. 
Localisation has been first applied to  gauge theories on spheres by  Pestun \cite{pestun},
 who obtained the partition function of  $\mathcal{N}=2$ theories on $S_4$. 
Kapustin, Willett, and Yaakov (KWY) \cite{KWY} applied localisation techniques to the study of $\mathcal{N}=2$ theories 
 on the three-sphere $S_3$.
Path integrals reduce  to  matrix models, which can be solved at large $N$.
  In particular, the ABJM matrix model, has been solved by 
 Drukker,  Mari\~no, and Putrov \cite{dmp}, who found the famous $N^{3/2}$ scaling of the entropy of multiple M2 branes.
Chern-Simons matter theories have also been studied \cite{tropical,kleb}.
For an excellent review on this topic and a complete list of references see \cite{mlec}.
Localisation techniques have been extended also to $\mathcal{N}=2$ theories
where the anomalous dimensions of the matter fields  are not canonical  \cite{jaffe,hoso}.

In \cite{kapmir} localisation has been applied to test non-perturbatively 
mirror symmetry in  strongly coupled super-conformal field theories in three dimensions deformed by real mass terms and FI parameters.
For conjectured mirror pairs of theories,  partition functions, computed by localisation,
 have been shown to agree provided the mass and FI parameters are exchanged.
In \cite{kapseib} Seiberg-like dualities have been tested with similar methods, while in \cite{ksv},
3d superconformal indices of  mirror pairs of theories  have been shown to coincide.
In this paper we apply the  KWY  localisation techniques to 3d $\mathcal{N}=4$ generalised quiver theories 
and to their star-shaped mirror dual.
One of our main results is the following  explicit expression for the partition function of the $T(SU(N))$ quiver tail Fig. \ref{tsun}:
$$
{\Z}^{T(SU(N))}(m_i; e_j)=\frac{\sum_{\rho  \in S^N} (-1)^\rho e^{2\pi i   \sum^N_j m_{\rho(j)} e_j}}
{ i^{N(N-1)/2}  \prod_{i<j}^N sh(m_i-m_j)  \prod_{i<j}^N sh(e_i-e_j)  },
$$
displaying a manifest self-mirror symmetry under the exchange of  mass  $m_i$ and FI $e_i$ parameters.
The $T(SU(N))$ quiver tail is the building block to construct generic star shaped quiver theories,
 by using our exact expression we are able to  compute  partition functions of this infinite family of theories
 solving exactly the matrix integrals at finite N.
For mirror pairs  involving only theories admitting a  Lagrangian description 
we compute partition functions on both sides of the duality and check that they
agree provided we exchanged mass and FI parameters.
In this way we provide infinite non-perturbative tests of the mirror construction of \cite{sici}.
Assuming mirror symmetry we then obtain the partition function of the $T_N$ theory which is the building block to construct generalised quiver gauge theories.

The papers is organised as follows.
In section \ref{loc} we introduce the KWY rules for the computation of  partition functions on $S_3$.
We then compute the partition function of the $U(1)$ theory with $N$ flavours which is one of our main tools.
In section \ref{n=2} we study several mirror pairs of  rank two models.
We begin section \ref{lagra} with the computation of the partition function of the $T(SU(N))$ theory.
We then use this building block to compute partition functions of star shaped quiver theories and compare them with their mirrors.
In section \ref{nonlagra} we study non-Lagrangian theories. We compute the partition function of the $T_N$ theory and use it as 
a building block to obtain generalised quiver theories.
We discuss the TQFT structure of these theories and the associativity of the $T_N$ blocks.

\section{Our set-up}\label{loc}

In this section we review the rules for the computation of  partition functions on $S_3$
and our main tool: the explicit result for the partition function of $U(1)$ theory with $N$ flavours.

\subsection{The Kapustin-Willet-Yaakov matrix integrals}

Recently it has been shown that the path integral of $3d$  supersymmetric theories localises to a matrix integral \cite{KWY}.
In the case of $\mathcal{N}=4$ quiver gauge theories, with $SU(N)/U(N)$ gauge groups and fundamental/bifundamental matter, the partition function on $S^3$ is given by a matrix integral that is written down using the following rules.\\
In order to have a slightly more concise notation, we define:
\ben
sh(A) \equiv 2 \sinh( \pi A), \qquad ch(A) \equiv 2 \cosh( \pi A).
\een
For every gauge group $U(N)$ with FI parameter $\eta$ we have the following integral over the Cartan divided by the residual Weyl symmetry:
\ben
\int_{-\infty}^{+\infty} \frac{\rd^Nx}{N!} \prod_{i<j}^N sh^2(x_i-x_j) e^{2 \pi i \eta \sum_i^N x_i}.
\een
For $SU(N)$  gauge groups we replace  $d^Nx$ with $\rd^Nx\delta(\sum x_i)$ and remove the FI parameter.

For every fundamental of mass $m$ attached to the node with integration variables $x_i$ we add the factor:
\ben
\frac{1}{\prod_{i=1}^N ch(x_i -m)},
\een
while for every bifundamental of mass $m$ attached to the nodes $U(N_1)$ with integration variables $x_i$ and $U(N_2)$ with integration variables $y_i$ we insert:
\ben
\frac{1}{\prod_{i=1}^{N_1}  \prod_{j=1}^{N_2} ch(x_i -y_j-m)}.
\een

\paragraph{Example}

\begin{figure}[!ht]
\leavevmode
\begin{center}
\label{unkf}
\includegraphics[height=1cm]{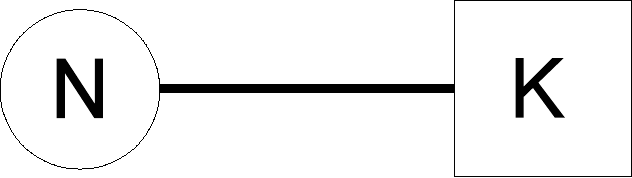}
\end{center}
\caption{ $U(N)$ with $K$ flavours.}
\label{nfree}
\end{figure}

To see our rules at work we consider the case of a  $U(N)$ theory with $K$ flavours with masses $m_j$ and  FI $\eta$, the corresponding quiver is depicted in Fig. \ref{unkf}.

The partition function reads:
\ben
\hat\Z^{U(N)}_{K} = \int \frac{\rd^Nx}{N!} \frac{\prod_{i<j}^N sh^2(x_i-x_j)}{\prod_{i=1}^N\prod_{j=1}^K ch(x_i-m_j)}  e^{2 \pi i \eta \sum_i^N x_i}. \een
Here and in the rest of this paper we  use the hat to indicate  off-shell 
partition functions. On-shell partition functions satisfy
$\sum_i^Km_i=0$.

\paragraph{The Cauchy determinant formula}
A tool that is very useful is the so called Cauchy determinant formula:
\ben
\frac{\prod_{i<j}^N sh(x_i-x_j) \cdot \prod_{i<j}^N sh(y_i-y_j)}{\prod^N_{i,j}ch(x_i-y_j)} = \sum_{\rho \in S^N}(-1)^{\rho}\frac{1}{\prod_{i=1}^N ch(x_i - y_{\rho(i)})}.
\een
This formula can be used to get rid of the  $\prod_{i<j}^N sh^2(x_i-x_j)$ associated to a certain node whenever the number of flavours for that node is twice the number of colours, or greater. That is when that node is a {\it good} node, in the Gaiotto-Witten sense \cite{GW}. For instance, in the case of $U(N)$ with $K \geq 2N$ flavours we can separate the flavors into 3 parts, $N+N+(K-2N)$ with masses $\{m_i\},\{\tm_i\},\{M_j\}$, and use the Cauchy determinant formula twice, to write $\Z^{U(N)}_{K} $ as
\ben \nonumber
\!\!\!\! \frac{\sum_{\rho, \rho'} (-1)^{\rho+ \rho'}}{\prod_{i<j}^N sh(m_i-m_j) sh(\tm_i-\tm_j)} 
\int  \frac{\rd^Nx}{N!}\frac{ e^{2 \pi i \eta \sum_i^N x_i}}{\prod_{i=i}^Nch(x_i-m_{\rho(i)})  ch(x_i-\tm_{\rho'(i)})\prod_{j=2N+1}^{K} ch(x_i-M_j) }.\\
\een
We see that  the integral factorizes into $N$ single (abelian) integrals and  in practice  we have to deal with
 a sum of products of partition functions of $U(1)$ gauge theories.
All the Lagrangian  $\mathcal{N}=4$ theories that we will study in this paper share this {\it abelianization} property, so, to compute  exact partition functions,
we will need the partition functions of  the $U(1)$ theory with $N$ flavours.

\subsection{Abelian  integrals: $U(1)$ with $N$ flavours}
In this section we compute the partition functions of  the $U(1)$ theory with $N$ flavours. 
\begin{figure}[!ht]
\leavevmode
\begin{center}
\includegraphics[height=1cm]{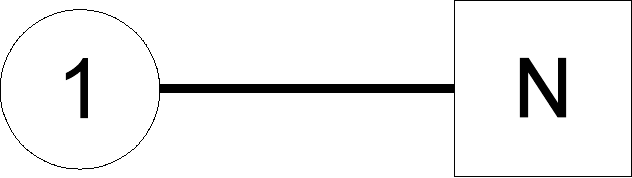}
\end{center}
\caption{The $U(1)$ theory with $N$ flavours.}
\label{db}
\end{figure} 

The partition function is defined as follows:
\ben
\hat{\Z}^{U(1)}_{N}(\eta;m_i) \equiv \int \rd s \frac{e^{2 \pi i s \eta}}{ch(s - m_1) ch(s -m_2 ) \cdots ch(s -m_N )}. 
\een
This is the off-shell partition function of a $U(1)$ gauge theory coupled to $N$ hypers of charge $1$ and masses $m_i$, which we assume to be real.
It is possible to perform the integral explicitly using residues, we have to consider $N$ infinite sets of simple poles, located at
\ben
s=m_i+ i (2k_i+1)/2  \qquad k_i=0,1,\ldots
\een
the first set contributes:
\ben
 \frac{e^{2 \pi i  m_1 \eta } }{\prod_{j \neq 1} ch(m_1-m_j+i/2) } e^{- \pi \eta}\sum^{\infty}_{k=0}  (-1)^{k N} e^{-2 \pi k\eta},
\een
taking into account all the sets of poles we obtain:
\ben\label{genericabel}
 \hat{\Z}^{U(1)}_{N}(\eta;m_i)&=& \frac{1}{(e^{\pi \eta}-(-1)^N e^{-\pi \eta} ) } \sum_{i=1}^N \frac{ e^{2 \pi i  m_i \eta} }{\prod_{j \neq i} ch(m_i-m_j+i/2)}=\nonumber\\
&=& \frac{1}{i^{N-1}(e^{\pi \eta}-(-1)^N e^{-\pi \eta} ) } \sum_{i=1}^N \frac{ e^{2 \pi i  m_i \eta} }{\prod_{j \neq i} sh(m_i-m_j)}\;.
  \een
In the limit of vanishing FI parameter $\eta \rightarrow 0$, 
we have:
\ben\label{genericabel0}
 \hat{\Z}^{U(1)}_{N}(0;m_i)=&0 &\; , \;\; odd \; N ,\\
 \hat{\Z}^{U(1)}_{N}(0;m_i)=& \frac{i}{ i^{N-1}} \sum_{i=1}^N \frac{m_i}{\prod_{j \neq i} sh(m_i-m_j)}&\; , \;\; even \; N.
\een

\subsubsection*{Example $N=1$:}  
In the special case of one flavour we get :  
\be\label{basic}
\hat{\Z}^{U(1)}_{1}(\eta, m) = \int \rd a \frac{e^{2\pi i a \eta } }{ch(a-m)}=\frac{e^{2 \pi i m \eta}}{ch(\eta)}.
\ee
On shell, for $m=0$ we have:
\be\label{basic}
\Z^{U(1)}_{1}(\eta) = \frac{1}{ch(\eta)} = \Z_{1 free H}(\eta),
\ee
which is a manifestation of the basic statement of abelian mirror symmetry \cite{is}:
the $\mathcal{N}=4$,  $U(1)$ theory with one flavour and with  FI parameter $\eta$, is mirror of the $\mathcal{N}=4$ theory of $1$ free hyper with mass $\eta$.



\subsubsection*{Example $N=2$:}  
In the special case of $2$ flavours we get:   
 \ben
\label{double}
\Z^{U(1)}_{2}(\eta, m) = \int \rd s \frac{e^{2 \pi i s \eta}}{ch( s - m/2 ) ch(s + m/2 ) } =\frac{-i\left( e^{ \pi i  m \eta } -  e^{ - \pi i m \eta} \right)}{sh(m/2+m/2)    s(\eta)  } =\frac{2 \sin( \pi \eta m)}{sh(m) sh(\eta)},
\een
we see that the partition function is symmetric in $m \leftrightarrow \eta$:
\be \Z^{U(1)}_{2}(\eta, m) = \Z^{U(1)}_{2}(m, \eta) .\ee
This is a manifestation of the fact that $U(1)$ with $2$ flavours is self-mirror \cite{is}.


\section{$U(2)$ and $SU(2)$ models}\label{n=2}  
In this section we compute partition functions of the following quiver gauge theories:
\begin{enumerate}
\item SU(2) with K flavours.
\item $U(2)^{k+1} \times U(1)^4 // U(1)$, the  $D_k$ quiver.
\item $SU(2) \times U(1)^N$ star-shaped.
\end{enumerate}
We will compute the partition functions of these models as functions of FI and mass parameters.
Some of these theories are related by mirror symmetry as explained in the following table
where we indicate the dimensions of  Higgs and Coulomb branches and the number of masses and FI parameters:

\vspace{.5cm}
\begin{tabular}{|c|c|c|c|c|}\hline
Model & dim Higgs & dim Coulomb & $\#$ masses & $\#$ of FI's \\ \hline\hline
SU(2) with K flavors & $2K-3$  & $1$ & $K$ & $0$\\ \hline
$SU(2) \times U(1)^N$ star-shape & $N-3$ & $N+1$ & $0$ & $N$ \\ \hline
$SU(2)^k$ linear-shape & $k+4$ & $k$ & $k+3$ & $0$ \\ \hline
$U(2)^{k+1} \times U(1)^4 // U(1)$ $D_k$-shape & $1$ & $2k+5$ & $0$ & $k+4$ \\ \hline
\end{tabular}

\vspace{.5cm}

One can easily  single out mirror pairs as pairs of theories for which the 
Higgs and Coulomb branches  are exchanged.

\subsection{$SU(2)$ with K flavours}

\begin{figure}[!ht]
\leavevmode
\begin{center}
\includegraphics[height=1cm]{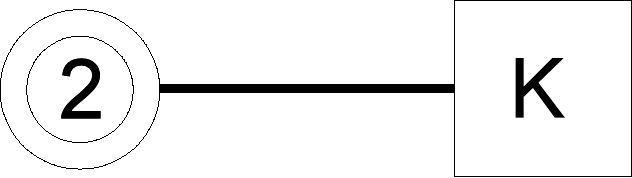}
\end{center}
\caption{ $SU(2)$ with $K$ flavours.}
\label{fsu2kf}
\end{figure}

The partition function with quiver diagram in Fig. \ref{fsu2kf}\footnote{We use a double circle, as opposed to a simple one to denote $SU(N)$ nodes.} 
is defined as
\ben \nn \hat \Z^{SU(2)}_{K}(m_i) =\int  \frac{\rd x_1 \rd x_2}{2!} \delta(x_1+x_2)   \frac{sh^2(x_1-x_2)}{\prod_{j=1}^K ch(x_1-m_j)ch(x_2-m_j)}. \een
We separate the $K \geq 4$ flavors into $2+2+(K-4)$ so we can use the Cauchy determinant formula twice and get rid of the $sh^2(x_1-x_2)$ numerator:
\ben 
\label{pfsu2kf}
&&\frac{1}{sh(m_1-m_2)sh(m_3-m_4)} \int \frac{\rd x}{\prod_{p=5}^{K} ch(x - m_p)ch(x + m_p)} \times \\
&&\nn \times \left( \frac{1}{ch(x-m_1)ch(x+m_2)}-\frac{1}{ch(x-m_2)ch(x+m_1)}\right) \frac{1}{ch(x-m_3)ch(x+m_4)}.
\een
Now we can perform the last integral over $x$ using  eq. (\ref{genericabel})  with $2(K-4)+4=2K-4$ flavours  and vanishing FI parameter.
  Let us focus on the terms proportional to $m_1$, there are $2$ such terms and they contribute as
\ben
\nn +\frac{m_1}{\prod_{i \geq 5} sh(m_1-m_i)sh(m_1+m_i) sh(m_1+m_2) sh(m_1-m_3)sh(m_1+m_4)}\\
\nn -\frac{-m_1}{\prod_{i \geq 5} sh(m_1-m_i)sh(m_1+m_i) sh(-m_1-m_2)sh(-m_1-m_3) sh(-m_1+m_4)},
\een
summing these two terms we get
\ben
\frac{m_1sh(2m_1)}{\prod_{j \neq 1}( sh^2(m_1) - sh^2(m_j))}.
\een
By symmetry in the $m_i$, we obtain the following exact expression for the partition function:
\ben
\label{esu2kf}
\hat \Z^{SU(2)}_{K}(m_i) = \sum_{i=1}^K \frac{m_ish(2m_i)}{\prod_{j \neq i}( sh^2(m_i) - sh^2(m_j))}.
\een

\subsection {The $D_k$ quiver}

In this section we study the $U(2)^{k+1} \times U(1)^4 // U(1)$ theory also known as the  $D_k$ quiver.
As in Fig. \ref{db} we  denote by  $\xi_I$   the  FI's and by  $z^{(I)}_i$ the Cartan's of the $U(2)$ nodes with $I=1,\cdots, k$, $i=1,2$.
We then denote by $x_i$ the Cartan's of the $SU(2)$ node 
and by $\eta_{a,b,c,d}$, $a,b,c,d$ the FI's and Cartan's of the $U(1)$ nodes.

\begin{figure}[!ht]
\leavevmode
\begin{center}
\includegraphics[height=5cm]{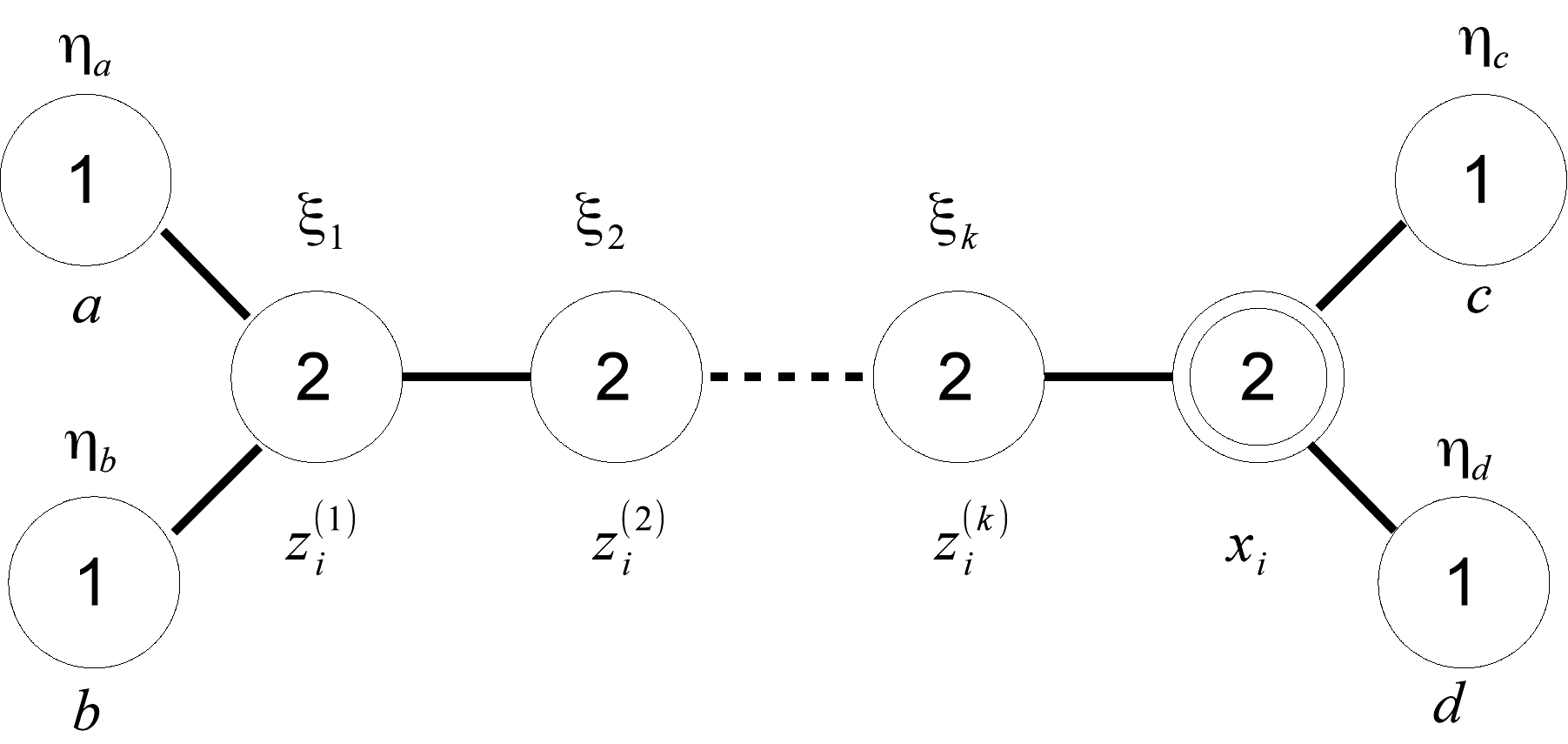}
\end{center}
\caption{the $D_k$ quiver}
\label{db}
\end{figure}

The partition function is given by:
\ben
\nonumber \mathcal{Z}_{D_k}&=&\frac{1}{2^{k+1} } \int  \rd a \rd b \rd c \rd d  \prod_I^k  \rd z^{I}_1 \rd z^{I}_2  \rd x_1 \rd x_2 \delta(x_1+x_2) \frac{   e^{2\pi i \sum_I \xi_I(z^{I}_1+z^{I}_2 )}   e^{2\pi i (\eta_a a +\eta_b b+\eta_c c +\eta_d d)}}{\prod_i^2 ch(z^{(1)}_i-a)  ch(z^{(1)}_i-b) ch(x_i-c) ch(x_i-d)     } 
\\ && \times \frac{sh(z^{(1)}_1-z^{(1)}_2)^2  sh(z^{(2)}_1-z^{(2)}_2)^2 \cdots sh(z^{(k)}_1-z^{(k)}_2)^2  sh(x_1-x_2)^2 }{
\prod_{i,j}^2 ch(z^{(2)}_i-z^{(1)}_j )\cdots
ch(z^{(k)}_i-z^{(k-1)}_j )ch(x_i-z^{(k)}_j ) }\nonumber =\\&&=\nonumber
\frac{1}{2^{k+1}}
 \frac{1}
 { sh\eta_a sh\eta_c}
 \int   \rd b  \rd d  \prod_I^k  \rd z^{I}_1 \rd z^{I}_2  \rd x_1 \rd x_2\delta(x_1+x_2)  \frac{   e^{2\pi i \sum_I \xi_I(z^{I}_1+z^{I}_2 )}   e^{2\pi i (\eta_b b +\eta_d d)}}{\prod_i^2   ch(z^{(1)}_i-b) ch(x_i-d)     } 
\\ && \times \nonumber
 \left(e^{2 \pi i \eta_a z^{(1)}_1} -  e^{2 \pi i \eta_a z^{(1)}_2} \right)
 \left(e^{2 \pi i \eta_c x_1} -  e^{2 \pi i \eta_c x_2} \right)
\\ && \times 
\frac{\sum_{\rho^{(1)}\cdots \rho^{(k)} } (-1)^{\rho^{(1)}+\cdots \rho^{(k)}}}{
\prod_{i}^2 ch(z^{(2)}_i-z^{(1)}_{\rho^{(1)}(i)} )\cdots
ch(z^{(k)}_i-z^{(k-1)}_{\rho^{(k-1)}(i)} )ch(x_i-z^{(k)}_{\rho^{(k)}(i)}  ) }.\een

By reordering the integration variables $z^{I}_i$ we get rid of $k$ sums over permutations and gain a  factor $2^k$.
It is now convenient to  take the (inverse) Fourier transform of each $ch$ to obtain:

\ben
\label{odk}
\nonumber\mathcal{Z}_{D_k}&=&\frac{1}{2}
 \frac{1}
 { sh\eta_a sh\eta_c}
 \int   \rd b  \rd d  \prod_I^k  dz^{I}_1 dz^{I}_2  \rd x_1 \rd x_2\delta(x_1+x_2)  e^{2\pi i \sum_I \xi_I(z^{I}_1+z^{I}_2 )}   e^{2\pi i (\eta_b b +\eta_d d)}
\\ && \times \nonumber
 \left(e^{2 \pi i \eta_a z^{(1)}_1} -  e^{2 \pi i \eta_a z^{(1)}_2} \right)
 \left(e^{2 \pi i \eta_c x_1} -  e^{2 \pi i \eta_c x_2} \right)\\
 \nonumber&&
\times \int \frac{\rd s_1 \rd s_2}{chs_1 chs_2}e^{2\pi i \left(s_1(z^{(1)}_1-b )+s_2(z^{(1)}_2-b )\right)   }
 \int \frac{\rd p_1 \rd p_2}{chp_1 chp_2}e^{2\pi i \left(p_1(x_1-d )+p_2(x_2-d )\right)   }
\\ && \times 
\int\frac{\prod_{I=2}^{k+1} \rd t^{(I)}_1 \rd t^{(I)}_2}{cht^{(I)}_1 cht^{(I)}_2}
e^{2\pi i \sum_{I=2}^k \left(t^{(I)}_1 (z^{(I)}_1-z^{(I-1)}_1 )+t^{(I)}_2(z^{(I)}_2-z^{(I-1)}_2 )\right)   }
e^{2\pi i \left(t^{(k+1)}_1 (x_1-z^{(k)}_1 )+t^{(k+1)}_2(x_2-z^{(k)}_2 )\right)   }.\nonumber\\
\een
We will now show that the above expression coincides  with  the partition function of the mirror theory,
 the $SU(2)$ theory with $K=3+k$ flavours (Fig. \ref{fsu2kf}),
provided we used the following  dictionary:
\ben
\nonumber\eta_a=m_4-m_3,\qquad  \eta_b=m_3+m_4,&&\qquad \eta_c=m_2-m_1,\qquad \eta_d=m_1+m_2\\
\xi_I= m_{4+I}-m_{3+I},&& I=1,\cdots k.
\een

There are four terms in eq. (\ref{odk}).
Let's consider first  the term proportional to   $e^{2 \pi i \eta_a z^{(1)}_1} e^{2 \pi i \eta_c x_1} $, the integration  over $b,d,x_i, z^{I}_i$ produces  the following deltas:
\ben
\nonumber &&\delta(s_1-t^{(1)}_1 +\xi_1+\eta_a), \qquad \delta(s_2-t^{(1)}_2 +\xi_1),\\ \nonumber
&&\delta(t^{(I-1)}_1 -t^{(I)}_1+\xi_I)\qquad \delta(t^{(I-1)}_2 -t^{(I)}_2+\xi_I ),\qquad I=2,\cdots k+1,\\
&&\delta(-s_1-s_2+\eta_b),\qquad \delta(-p_1-p_2+\eta_d)\nonumber,\\
&&\delta(p_1-p_2+t^{(k+1)}_1-t^{(k+1)}_2+\eta_c),\nonumber\\
\een
solving for $s_1$ and expressing the result in  terms of 
$x=s_1+m_3$ we obtain the following combination of $ch's$ in the denominator:
\ben
\label{one}
&&
ch(x+m_3) ch(x-m_4)ch(x+m_5)ch(x-m_5)\cdots
ch(x+m_{k+4})ch(x-m_{k+4})\nonumber\\ &&ch(x+m_2)ch(x-m_1).
\een
The remaining three terms give similar contributions.
Putting everything together   we rewrite the partition function as:
\ben\nonumber
\mathcal{Z}_{D_k}&=&\frac{1}{sh(m_2-m_1)sh(m_4-m_3)} \int \rd x \frac{1}{\prod_I^{k} ch(x+m_{4+I})ch(x-m_{4+I})  }\\ \nonumber&&\times\left(\frac{1}{ch(x-m_1) ch(x+m_2)}-\frac{1}{ch(x+m_1) ch(x-m_2)}\right)\frac{1}{ch(x-m_3) ch(x+m_4)}.\\
\een

This expression coincides with the partition function of the $SU(2)$ theory with $K=3+k$ flavours, eq. (\ref{pfsu2kf}).

 \subsection{$SU(2)\times U(1)^N$  star shaped} 
\begin{figure}[!ht]
\leavevmode
\begin{center}
\includegraphics[height=3cm]{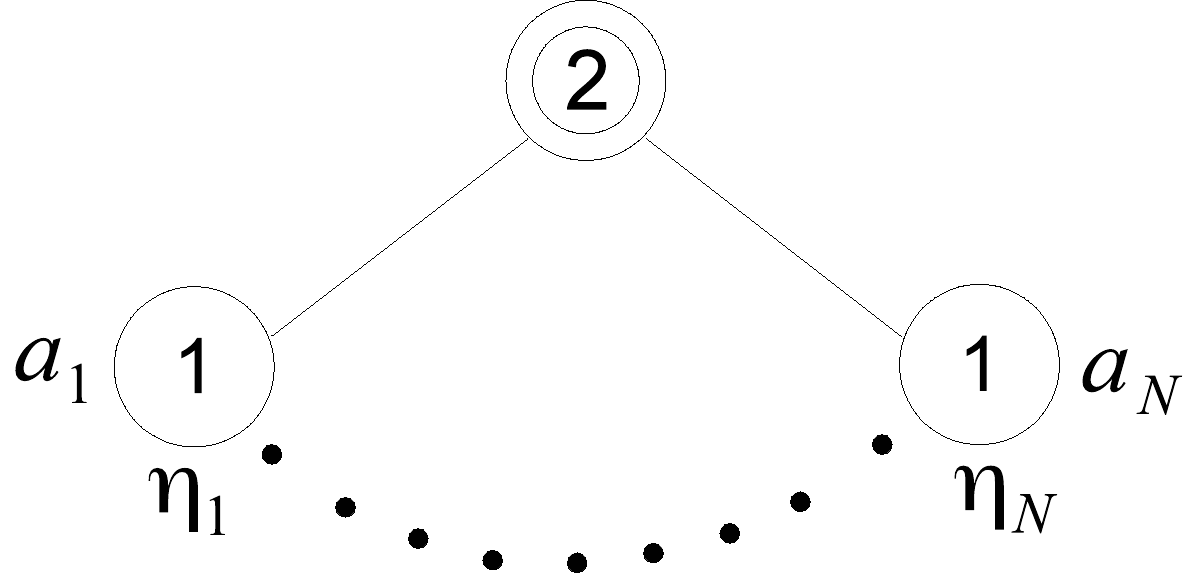}
\end{center}
\caption{ $SU(2)\times U(1)^N$.}
\label{su2nu1}
\end{figure}

 Let's consider now the  $SU(2)\times U(1)^N$ star shaped theory depicted in Fig. \ref{su2nu1},
  with FI parameters $\eta_i$, $i=1,\cdots N$. 

The partition function reads:
 \ben
 \Z_N(\eta_i)=\int \prod_i^N da_i      \frac{\rd x_1 \rd x_2}{2!}  \delta(x_1+x_2)e^{2 \pi i \sum_{i} \eta_i a_i} \frac{sh^2(x_1-x_2) }{
\prod_{i=1}^2\prod_{j=1}^N   ch(x_i-a_j)}.
\een
The Abelian integrals over $a_i$ can be performed using the result in  (\ref{double}) and we obtain:
\ben  \label{genstar}
 \Z_N(\eta_i)&\nonumber=&\frac{1}{2} \int    \rd x_1 \rd x_2  \delta(x_1+x_2) \prod_i \hat{\Z}^{U(1)}_{2}(\eta_i; x_1,x_2) \\
&=& \frac{1}{2}\int    \frac{\rd x}{sh(2x)^{N-2}}  \prod_{i=1}^N  \frac{2  \sin(2\pi \eta_i x ) }{sh(\eta_i)}.
\een
The  integral is convergent for $N \geq 3$.
For $N<3$ we have  {\it bad} theories in the Gaiotto-Witten \cite{GW} sense while $N=3$ corresponds to an {\it ugly} theory mirror of the theory of $4$ free hypers. For $N>3$ the mirror is a linear quiver with $N-3$ SU(2) gauge groups, a {\it good} theory. In  eq. (\ref{genstar}) the $S_N$ symmetry among the $N$ FI parameters is explicit, and the last integral can be performed in terms of the $\sin/\cos$ transform of $sinh(t)^{-k}$, after expanding the product of the $N$ $\sin(\eta x)$'s in terms of sums of $\sin/\cos$ functions. We will perform the integral for the cases $N=3$ and $N=4$.

More results for generic $SU(2)$ star shaped quivers can be obtained as particular cases 
of the results  we  give in section \ref{stars}.

\subsubsection*{Example $SU(2)\times U(1)^3$:}  

In order to perform the computation of $\Z_3$ we use the trigonometric identity
$$2\sin(a)2\sin(b)2\sin(c)=2\left(\sin(a+b+c)+\sin(a-b-c)+\sin(-a+b-c)+\sin(-a-b+c)\right)$$
and  the $sin$-transform:
$$ \int \rd s \frac{\sin(2\pi \eta s)}{sh(2 s)}=\frac{ sh(\eta/2)}{2 ch(\eta/2)},$$
 to get
\ben \nonumber
\Z_3= \frac{th((\eta_1+\eta_2+\eta_3)/2)+th((\eta_1-\eta_2-\eta_3)/2)+th((-\eta_1+\eta_2-\eta_3)/2)+th((-\eta_1-\eta_2+\eta_3)/2)}{2sh(\eta_1)sh(\eta_2)sh(\eta_3)}.
\een
After few manipulation  the above expression simplifies to\footnote{This expression has been previously  obtained in \cite{gadde}.}:
\ben
\nonumber
\Z_3= \frac{1}{2ch((\eta_1+\eta_2+\eta_3)/2)ch((\eta_1-\eta_2-\eta_3)/2)ch((-\eta_1+\eta_2-\eta_3)/2)ch((-\eta_1-\eta_2+\eta_3)/2)}.\\\een
This is the partition function of $4$ free hypers with masses $(\eta_1 \pm \eta_2 \pm \eta_3)/2.$

\subsubsection*{Example $SU(2)\times U(1)^4$:} 

 In the special case $N=4$ we need to use a trigonometric identity to express the product of $4$ $\sin(x)$'s in terms of the sum of $8$ $\cos(x)$'s:
\ben
\label{z4}
\nonumber \frac{1}{2}&&\!\!\!\!\!\!\int  \frac{ \rd x}{sh(2x)^{2}}   \prod_{j=1}^4 2 sin(2\pi \eta_i x )=\int  \frac{\rd x}{sh(2x)^{2}}
 \big(\cos(2\pi( \eta_1+\eta_2+\eta_3+\eta_4 )x)+
\cos(2\pi( \eta_1+\eta_2-\eta_3-\eta_4 )x) +\\&&\!\!\!\!\!\! \nonumber+\cos(2\pi( -\eta_1+\eta_2+\eta_3-\eta_4 )x) +
\cos(2\pi( \eta_1-\eta_2+\eta_3-\eta_4 )x)-\cos(2\pi(- \eta_1+\eta_2+\eta_3+\eta_4 )x)-\\ \nonumber &&\!\!\!\!\!\! -\cos(2\pi( \eta_1-\eta_2+\eta_3+\eta_4 )x)
-\cos(2\pi( \eta_1+\eta_2-\eta_3+\eta_4 )x)-\cos(2\pi( \eta_1+\eta_2+\eta_3-\eta_4 ) x)\big)\\
\een
and use the following $\cos$-transform:
\ben \nonumber
\int \rd x    \frac{ \cos(2\pi \xi x )}{sh(2x)^{2}}=\frac{\xi ch(\xi/2)}{2 sh( \xi/2)} 
\een
to compute the last integral.

At this point we change variables with  the dictionary:
\ben
\eta_{1}=\tilde m_1-\tilde m_2,\qquad \eta_{2}=\tilde m_1+\tilde m_2,\qquad \eta_{3}= m_1- m_2,\qquad \eta_{4}= m_1+ m_2.  
\een  

We now collects  terms  proportional to $m_1$:
\ben
\frac{m_1\left( 
\frac{ch(m_1+\tm_1)}{sh(m_1+\tm_1)}+\frac{ch(m_1-\tm_1)}{sh(m_1-\tm_1)} -\frac{ch(m_1-\tm_2)}{sh(m_1-\tm_2)}-\frac{ch(m_1+\tm_2)}{sh(m_1+\tm_2)}
\right)}{sh(m_1-m_2)sh(m_1+m_2)sh(\tm_1-\tm_2)sh(\tm_1+\tm_2)}.
\een
We get similar expressions for the other $m$'s.
After few manipulations, putting all together we obtain\footnote{Set $m_3=\tilde m_1$ and $m_4=\tilde m_2$.}:

 \ben
\Z^{SU(2)\times U(1)^4}(m_i) = \sum_{i=1}^4  \frac{m_i sh(2 m_i)}{\prod_{j \neq i} \left( sh^2(m_i)-sh^2(m_j) \right)}.
\een
This is precisely the partition function of the $SU(2)$ theory  with $4$ flavours given in eq. (\ref{esu2kf}).


\section{Lagrangian theories}\label{lagra}

We will now move to the study of generalised quiver theories corresponding to spheres with two  generic punctures and any number of simple ones.
These theories admit a Lagrangian description. We will compute explicitly their partition functions and those of their mirror pairs, which are star shaped quiver theories.
We start with the explicit evaluation of the partition function of the $T(SU(N))$ quiver theory  which, being the mirror of a full puncture, is the main building block.

\subsection{$T(SU(N))$}

\begin{figure}[!ht]
\leavevmode
\begin{center}
\includegraphics[height=3cm]{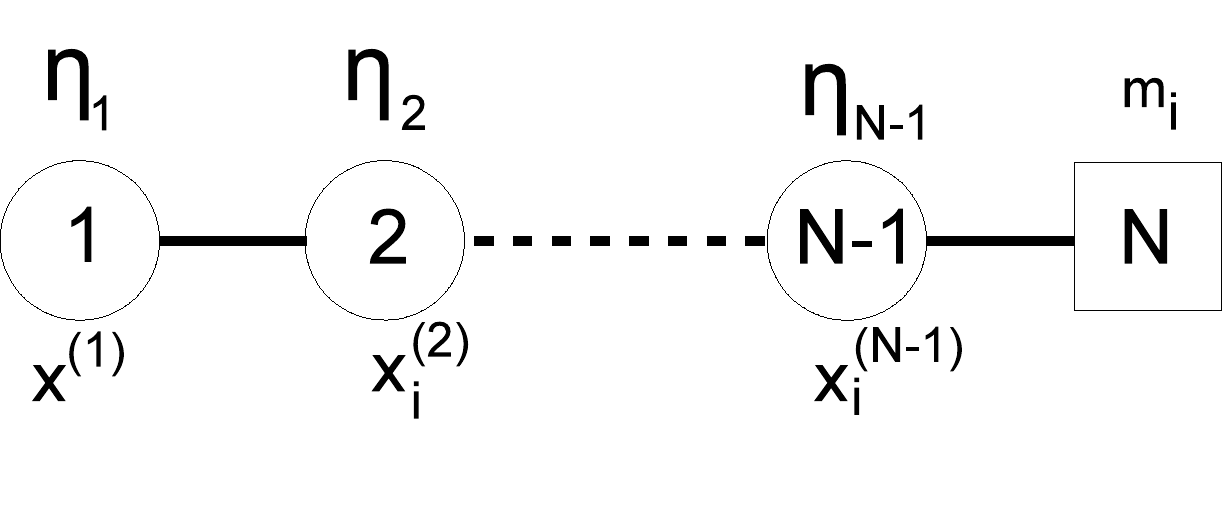}
\end{center}
\caption{The quiver diagram for $T(SU(N))$}
\label{tsun}
\end{figure}

In this section we will compute the partition function of  the $T(SU(N))$ quiver theory depicted in Fig. \ref{tsun}.
$T(G)$  is a $3d$ $\mathcal{N}=4$ gauge theory
at the IR super-conformal fixed point,  with global symmetry $G\times G^L$ ($G^L$ is the Langlands dual of $G$).
 The Higgs and Coulomb branches are respectively acted by $G$ and $G^L$.
Under the $\mathcal{N}=4$  mirror transformation, $T(G)$ is mapped to $T(G^L)$.
In the $G= SU(N)$ case the Coulomb and Higgs branches  are isomorphic and FI and mass parameters are exchanged by mirror symmetry.

$T(SU(N))$  will be our  fundamental building block to
compute the partition function of generic star shaped quivers corresponding to spheres with generic punctures.
Indeed, in \cite{sici}  the quiver tail $T(SU(N))$ has been identified with   the mirror of the full puncture $\odot$, we will then equivalently denote the $T(SU(N))$ partition function as $\mathcal{ Z}^{T(SU(N))}$ or as $\Z^{\odot}_N$.

Let's fix the notation as in Fig. \ref{tsun}.
Let $\eta_d$, $d=1,\cdots N-1$ be the FI parameters  and $x^{(d)}_i $ (with $i=1,\cdots d$) the Cartan's of the  $U(1)\times\cdots\times U(N-1)$ nodes. Let also $m_i$, $i=1,\cdots N$, $\sum_i^N m_i=0$ be the masses acted by the $SU(N)$ flavour symmetry.
The off-shell $T(SU(N))$ partition function is given, by the KWY rules, by:
\ben \label{DEFTSUN}
\mathcal{\hat Z}^{T(SU(N))}(m_1,\ldots, m_N;\eta_1,\ldots, \eta_{N-1})&=&\int \rd x^{(1)} \times \frac{\rd x^{(2)}_1 \rd x^{(2)}_2}{2!} \cdots \times \prod_i^{N-1} \frac{\rd x^{(N-1)}_i}{(N-1)!}
\times \\&&
\!\!\!\!\!\!\!\!\!\!\!\!\!\!\!\!\!\!\!\!\!\!\!\!\!\!\!\!\!\!\!\!\!\!\!\! \times \nonumber  e^{2\pi i \left( \eta_1 x^{(1)}+\eta_2 (x^{(2)}_1+x^{(2)}_2       ) +\cdots +\eta_{N-1} ( x^{(N-1)}_1+\cdots+x^{(N-1)}_{N-1}       ) \right)}\times \nonumber\\&&\!\!\!\!\!\!\!\!\!\!\!\!\!\! \!\!\!\!\!\!\!\!\!\!\!\!\!\! \!\!\!\!\!\!\!\!\!\!\!\!\!\! 
\!\!\!\!\!\!\!\!\!\! \!\!\!\!\!\!\!\!\!\!\!\!\!\!\!\!\!\!\!\!\!\!\!\!\!\!\!\!\!\!\!\!\!\!\!\! \times \nonumber \frac{
\prod_{d=2}^{N-1 }\prod_{i<j}^d sh^2(x^{(d)}_i-x^{(d)}_j )}{
\prod_{i=1}^N  \prod_j^{N-1} ch(m_i-x^{(N-1)}_j ) \cdots
\prod_i^{d}\prod_{j=1}^{d-1} ch(x^{(d)}_i-x^{(d-1)}_j ) \cdots
\prod_{j=1}^2 ch(x^{(2)}_j-x^{(1)} )
}.\een

It is convenient to change variables for the FI parameters, from $\eta_i$ to $e_i$:
\ben
\eta_i = e_i - e_{i+1}, \qquad {\rm for } \quad i=1,\ldots, N-1\qquad  {\rm with}\qquad \sum^N_ie_i=0.
\een

%
We claim that the result of the integral (\ref{DEFTSUN}) is:
\ben \label{TSUNMAIN}
\hat{\Z}^{T(SU(N))}(m_i; e_j)=\frac{\sum_{\rho  \in S^N} (-1)^\rho e^{2\pi i   \sum^N_j m_{\rho(j)} (e_j-e_N)}}
{ i^{N(N-1)/2}  \prod_{i<j}^N sh(m_i-m_j)  \prod_{i<j}^N sh(e_i-e_j)  }.
\een
This is one of our main results. We will provide a proof of this formula by induction at the end of this section.

In terms of the variables $\eta_i, i=1,\ldots,N-1$, the formula is a bit more complicated:


\ben
\hat{\Z}^{T(SU(N))}(m_i; \eta_i)=\frac{\sum_{\rho \in S^N} (-1)^\rho e^{2\pi i   \sum_{i=1}^{N-1} m_{\rho(i)} (\eta_{N-1}+ \cdots+\eta_i)}}{ i^{N(N-1)/2}  \prod_{i<j}^N sh(m_i-m_j)    \prod_{d=1}^{N-1}\prod_{k=1}^d  sh(\eta_d+\eta_{d-1}+\cdots +\eta_k)}.
\een

Let us check what happens for $N=2$, where $T(SU(2))$ is simply the $U(1)$ theory with $2$ flavours.
In this case eq. (\ref{TSUNMAIN}) reduces as expected to:
\ben
\hat{\Z}^{T(SU(2))}(m_1, m_2; \eta)=\frac{ e^{2\pi i   m_1 \eta }-e^{2\pi i   m_2 \eta}}{ i   sh(m_1-m_2)  sh(\eta)  } = \hat{Z}^{U(1)}_{2}(m_1, m_2; \eta).
\een

\paragraph{Check of the self-mirror property:}
The expected self-mirror property  of  the $T(SU(N))$ theory, which  exchanges $m_i\leftrightarrow e_i$,
 is manifest in our explicit expression for the partition function
 eq. (\ref{TSUNMAIN}).  By using that:
\be \sum_{\rho \in S^N} (-1)^\rho e^{2\pi i   \sum_j m_{\rho(j)}e_j}=\sum_{\rho' \in S^N} (-1)^{\rho'} e^{2\pi i   \sum_j e_{\rho'(j)} m_j}, \ee
it is clear that the on-shell partition function ($\sum_i^N m_i=0$) is self-mirror:
\be {\Z}^{T(SU(N))}(m_i; e_j) = {\Z}^{T(SU(N))}(e_i; m_j). \ee

\subsubsection{Proof of the formula by induction}
To prove our expression for $ \hat{\Z}^{T(SU(N))}$ we use it as a building block to construct  $\hat{\Z}^{T(SU(N+1))}$;
we gauge the flavour symmetry multiplying  by the $\prod^N_{i<j} sh^2(x_i-x_j) $ and integrating over the $SU(N)$ Cartan.
We then add the  FI  parameter $\eta_N$ and $N+1$ fundamentals
of masses $m_i$, $i=1, \cdots, N+1$:
\ben
\hat{\Z}^{T(SU(N+1))}= \int \frac{\rd^Nx}{N!}\prod^N_{i<j} sh^2(x_i-x_j)  \frac{\hat{\Z}^{T(SU(N))}(x;\eta_1\ldots\eta_{N-1})}{\prod^N_{i=1}\prod_{j=1}^{N+1} ch(x_i-m_j)}  e^{2 \pi i \eta_N \sum^N_i x_i}.
\een
Now we separate the $N+1$ fundamentals in a group of $N$ masses $m_i$, $i=1,\cdots N$ plus a singlet $m_{N+1}$ and plug our result for $\hat{\Z}^{T(SU(N)}$ to get:
\ben
\nonumber \hat{\Z}^{T(SU(N+1))}= &&\frac{1}{\prod_{i<j}^N sh(e_i-e_j)} \int \frac{\rd^Nx}{N!} e^{-2 \pi i e_N \sum^N_k x_k} \sum_{\rho \in S^N} (-1)^\rho e^{2\pi i   \sum^N_j e_{\rho(j)} x_j} e^{2 \pi i \eta_N \sum^N_i x_i}\\ && \times\frac{\prod^N_{i<j} sh(x_i-x_j) }{ i^{N(N-1)/2}     \prod^N_{i,j=1}ch(x_i-m_j)   \prod^N_{i=1} ch(x_i-m_{N+1})  },
\een
we now  use the Cauchy determinant formula once
\ben
\!\!\!\!\!\nonumber \frac{1}{ \prod_{p<q}^N sh(m_p-m_q)  \prod_{i<j}^N sh(e_i-e_j)}\sum_{\rho, \rho' \in S^N}(-1)^{\rho+\rho'} \int \frac{\rd^Nx}{N!}   \frac{ e^{-2 \pi i e_N \sum^N_i x_i} e^{2\pi i   \sum^N_j e_{\rho(j)} x_j} e^{2 \pi i \eta_N \sum^N_i x_i}}{  i^{N(N-1)/2}   \prod^N_{i} ch(x_i-m_{N+1})ch(x_i-m_{\rho'(i)})},\\
\een
we then reorder the integration  variables $x_i$ we get rid of one sum over permutations and cancel the  factor $N!$ and get:
\ben
\frac{1}{ \prod_{p<q}^N sh(m_p-m_q)  \prod_{i<j}^N sh(e_i-e_j)}\sum_{\rho\in S^N}(-1)^{\rho}
 \int  d^Nx   \frac{  e^{ 2\pi i   \sum^N_j x_j (e_j-e_{N+1})   }}{i^{N(N-1)/2}\prod^N_{i} ch(x_i-m_{N+1}) ch(x_i-m_{\rho(i)}) }, \nonumber\\
 \een
where we introduced the new variable $e_{N+1}$ by $\eta_N\equiv e_N-e_{N+1}$ . The integral at this point factorizes into abelian $2$-flavors integrals:
\ben
\label{proof}
&&\!\!\!\!\!\!\!\frac{1}{i^{N(N-1)/2} \prod_{p<q}^N sh(m_p-m_q)  \prod_{i<j}^N sh(e_i-e_j)}\sum_{\rho\in S^N}(-1)^{\rho}
 \prod^N_{i=1}  \frac{ \left( e^{ 2\pi i   m_{\rho(i)} (e_j -e_{N+1})   }       - e^{ 2\pi i   m_{N+1} (e_i -e_{N+1})   }       \right)}{  i sh(m_{N+1}-m_{\rho(i)}) sh(e_i-e_{N+1}) }  \nonumber \\
 &&=\frac{1}{i^{N(N+1)/2}   \prod_{p<q}^{N+1} sh(m_p-m_q)  \prod_{i<j}^{N+1} sh(e_i-e_j)}
 \sum_{\rho' \in S^{N+1} }       (-1)^{\rho'}   e^{2\pi i \sum_{i=1}^N  m_{\rho'(i)}(e_i   -e_{N+1}    )   } ,\een
 where in the last line  we made use of the following identity\footnote{The validity of this formula can be seen as follows. Separate all the $N!\cdot 2^N$ terms on the L.H.S. according to the number of times they contain the factor $e^{m_{N+1}}$. If $K>1$, all the terms containing $m_{N+1}$ precisely $K$ times cancel out among themselves when performing the sum over the $S^N$-permutations $\rho$, due to the $(-1)^{\rho}$ prefactor. So we are left with $N!$ terms that do not contain $m_{N+1}$ and $N! \cdot N$ terms that contain $m_{N+1}$ exactly once, which is precisely the content of the R.H.S.}:
 \ben
\sum_{\rho \in S^N}(-1)^\rho \prod_{i=1}^N \left(  e^{m_{\rho (i)} \alpha_i  } -  e^{m_{N+1} \alpha_i } \right)=
\sum_{\rho' \in S^{N+1} }       (-1)^{\rho'}    \prod_{i=1}^N e^{ m_{\rho'(i)}\alpha_i } .
\een
 
The last line in eq. (\ref{proof}) is precisely $\hat{\Z}^{T(SU(N+1)}$, this concludes our proof.

\subsection{Two maximal and one minimal puncture}\label{sfree}

In this section we will study the theory on  a sphere with two maximal and one minimal puncture.
This theory is {\it ugly} in the Gaiotto-Witten sense and corresponds to $N^2$ free hypers.
We will evaluate the partition function of the mirror theory the star shaped quiver obtained by gluing two full punctures ${\Z}_N^{\odot}$ and one abelian integral $ \hat{\Z}^{U(1)}_N$, corresponding to a simple puncture ${\Z}^{\times}$, see Fig. \ref{nfree}.

\begin{figure}[!ht]
\leavevmode
\begin{center}
\includegraphics[height=4cm]{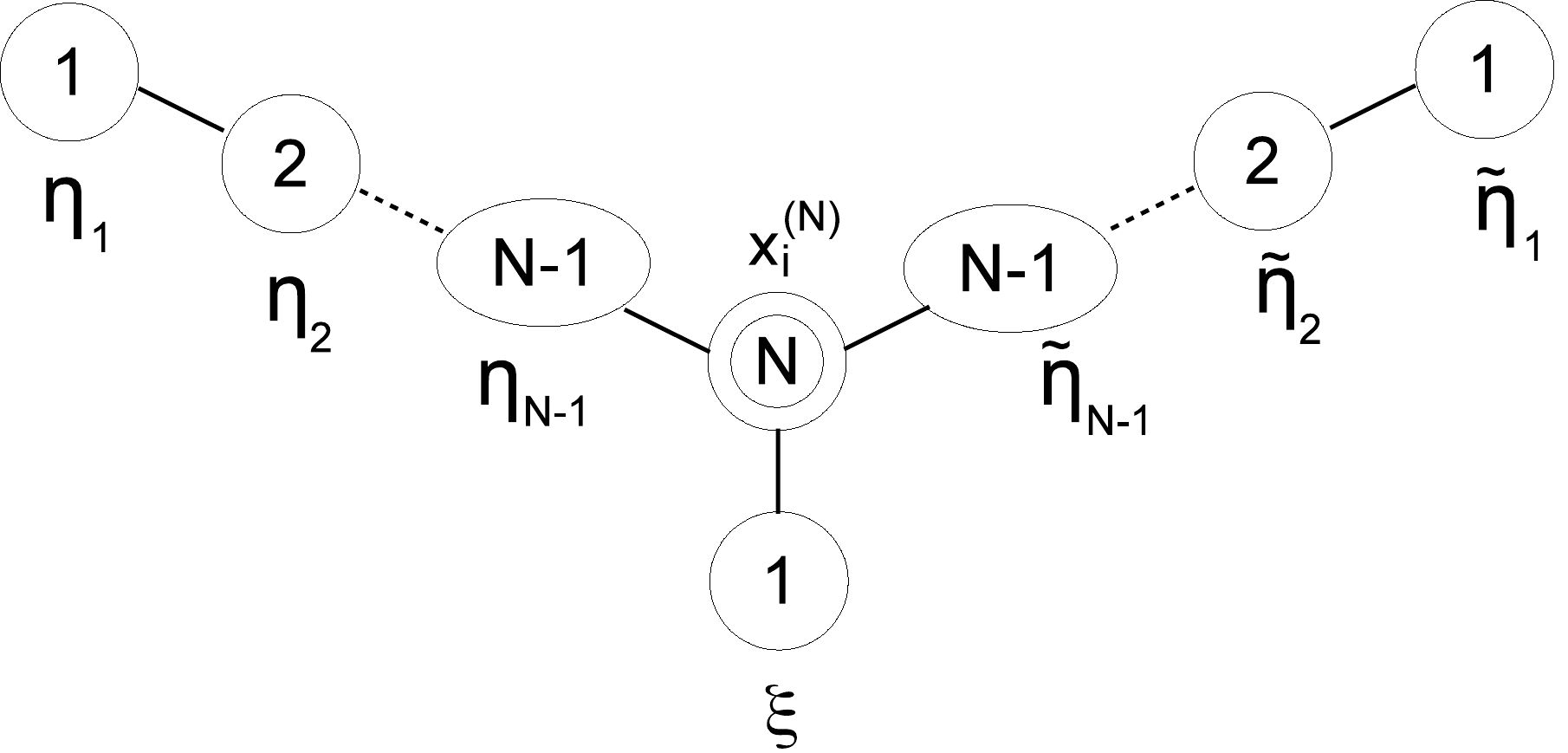}
\end{center}
\caption{Two maximal and one minimal punctures}
\label{nfree}
\end{figure} 
The partition function reads:
\ben
\nonumber \Z^{\times \odot \odot}_N(\xi,\tilde e,e)=\int \frac{\rd a \rd^Nx \delta(\sum x)\prod_{m<n}sh^2(x_m-x_n)}{N!}  \frac{e^{2 \pi i \xi a}}{\prod_i^N ch(x_i-a)} {\Z}^{\odot}_N(x; e_i){\Z}^{\odot}_N(x; \tilde{e_i}),\\
\een
by plugging in the result for $\Z^{T(SU(N))}$ we obtain:
\ben
\label{if}
&&\!\!\!\!\!\!\!\!\frac{1}{i^{N(N-1)} \prod_{i<j}^N sh(e_i-e_j) \prod_{i<j}^N sh(\te_i-\te_j)}\sum_{\rho,\tilde{\rho}} (-1)^{\rho+\tilde{\rho}}  \int \frac{\rd a \rd^Nx \delta(\sum x) e^{2 \pi i  \xi a} e^{2 \pi i \sum_j x_j (e_{\rho(j)} +\te_{\tilde{\rho}(j)}   )}}{N!\prod_i^N ch(x_i-a) } \nonumber\\&&=
\frac{1}{i^{N(N-1)} \prod_{i<j}^N sh(e_i-e_j) \prod_{i<j}^N sh(\te_i-\te_j)}\sum_{\rho} (-1)^{\rho}  \int \frac{\rd a \rd^N x \delta(\sum x) e^{2 \pi i  \xi a} e^{2 \pi i \sum_j x_j (e_{\rho(j)}+\te_j   )}}{\prod_i^N ch(x_i-a) },\nonumber\\
\een
where we  changed integration  variable to remove one sum over permutations.
We need the following integral with $A=(A_1, A_2, \ldots, A_N)=(\{e_{\rho(j)}+\te_j\})$:
\ben 
&&\nonumber  \int \rd a \rd^Nx \delta(\sum x)  \frac{e^{2 \pi i (\xi a + \sum_j x_j A_j)}}{\prod_i^N ch(x_i-a) } =   \int \rd a \rd^Nx \delta(\sum x+N a)  \frac{e^{2 \pi i ((\xi+\sum_i A_i) a + \sum_j x_j A_j)}}{\prod_i^N ch(x_i) }= \\
&&\nonumber = \frac{1}{N}  \int \rd^Nx   \frac{e^{2 \pi i \sum_j x_j (A_j-(\xi+\sum A_i)/N)}}{\prod_i^N ch(x_i) }= \frac{1}{N \prod_i^N ch(\xi/N - \langle A,h_i\rangle)},
\een
where  $h_i$ are the weights of $SU(N)$:
\ben (h_j)^I = \delta^I_j - \frac{1}{N}. \een
The partition function becomes
\ben
\Z^{\times \odot \odot}_N(\xi,\tilde \eta,\eta)&=&\frac{1}{ i^{N(N-1)}\prod_{i<j}^N sh(e_i-e_j) \prod_{i<j}^N sh(\te_i-\te_j)}\nonumber \times \\ &&\times \sum_{\rho} (-1)^{\rho}  \frac{1}{N \prod_i^N ch(\xi/N - \langle e_{\rho(i)},h_i\rangle- \langle\te_i,h_i\rangle)}.\nonumber\\
 \een
Finally we used the Cauchy determinant formula to get:
\ben
\label{free}
\Z^{\times \odot \odot}_N(\xi,\tilde \eta,\eta) &=& \frac{1}{i^{N(N-1)} N }  \prod_{i,j}^N
 \frac{1}{  ch(\xi/N - \langle e_i,h_i\rangle- \langle \te_j,h_j\rangle)},
 \een
which is, up to a prefactor, the partition function of $N^2$ free hypers, as expected from mirror symmetry.


\subsection{Two maximal and $k+2$ minimal}\label{stars}

\begin{figure}[!ht]
\leavevmode
\begin{center}
\includegraphics[height=2.5cm]{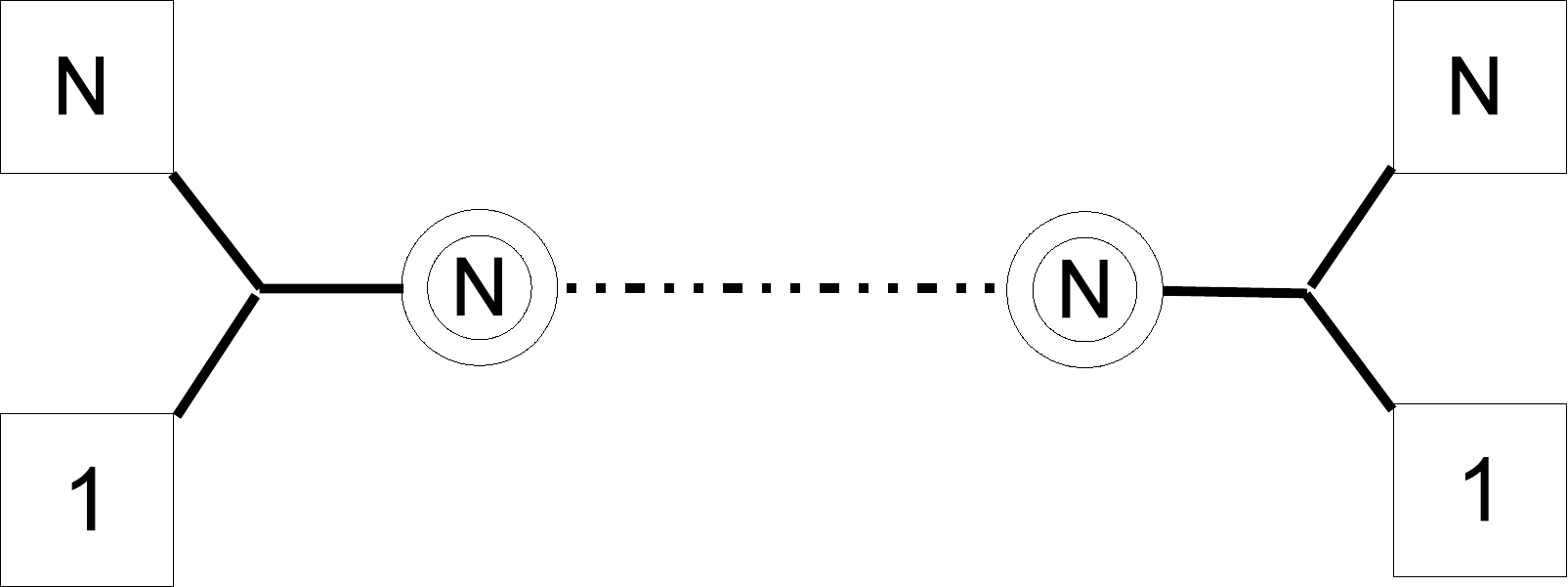}
\end{center}
\caption{ $SU(N)^K$ quiver.}
\label{ksun}
\end{figure}

In this section we study an infinite family of mirror theories associated to the sphere with two maximal and $k$ minimal punctures.
On one side we have the linear quiver theories $SU(N)^{k+1}$ with $N+N$ fundamentals $m_i,\tilde m_i$, $i=1,\cdots N$ acted by a $U(N)^2$ flavour symmetry, and $k$ bi-fundamentals $M_j$, $j=1,\cdots k$ depicted in Fig. \ref{ksun}.

\begin{figure}[!ht]
\leavevmode
\begin{center}
\includegraphics[height=4cm]{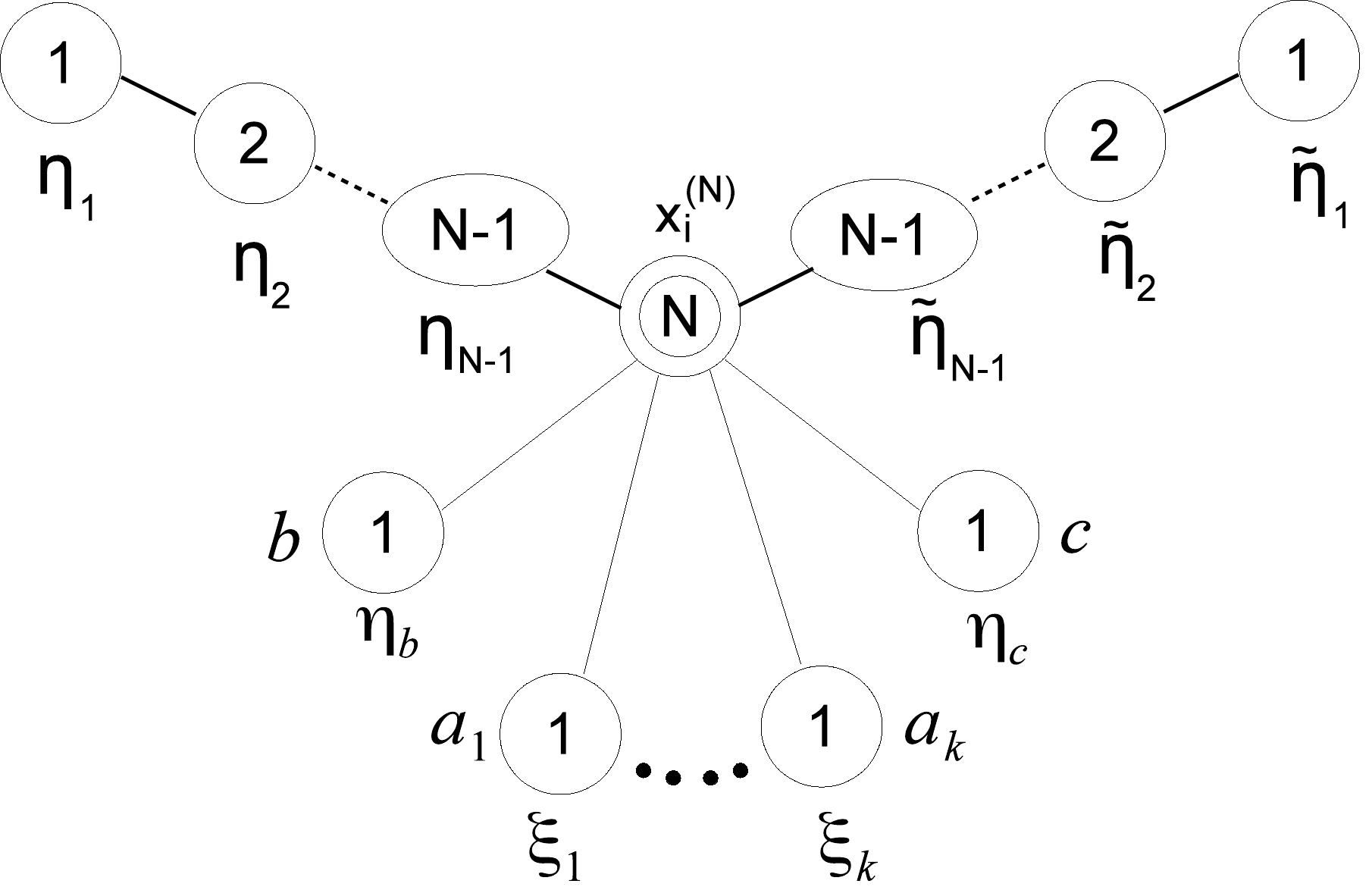}
\end{center}
\caption{Mirror of two maximal and $k+2$ minimal punctures.}
\label{codine}
\end{figure}

On the other side we have  the  the star shaped quivers 
$Z^{\times\cdots \times  \odot \odot}$, obtained by gluing two  full punctures  ${\Z}^{\odot}$ and $k+2$ abelian integrals $ \hat{\Z}^{U(1)}_N$ with FI's  $\eta_b,\eta_c,\xi_1\cdots \xi_k$ and Cartan's $b,c,a_1\cdots a_k$ as indicated in Fig. \ref{codine}.

We start from the linear quiver theory.

The partition function reads:
\ben
\nonumber\Z^{SU(N)^{k+1}}_{2N}&=&\frac{1}{(N!)^{k+1}}\int \prod_i^N \rd x_i \rd z_i \prod_r^{k-1} \rd y^{(r)}_i \delta(\sum_i x_i)\delta(\sum_i z_i)\delta(\sum_i y^{r}_i)\\ &&\nonumber  
\!\!\!\!\!\!\!\!\!\!\!\!\times \frac{\prod_{i<j}^N sh(x_i-x_j)^2 \prod_r^{k-1} sh(y^{(r)}_i -y^{(r)}_j )^2  sh(z_i-z_j)^2}{ch(x_i-m_j) ch(y^{(1)}_i-x_j-M_1 )ch(y^{(2)}_i-y^{(1)}_j-M_2 )
\cdots ch(z_i-y^{(k-1)}_j-M_k)  ch(z_i-\tilde m_j)}.
\een
We now use the Cauchy determinant $k+1$ times and  get $k+2$ sums over permutations of $S_N$. By changing variables it is possible to reorder $k+1$
 permutations and we find:
 \ben 
\nonumber\frac{1}{\prod^N_{i<j}sh(m_i-m_j)sh(\tm_i-\tm_j)}\sum_{\rho \in S^N}(-)^{\rho}
\int \prod_i^N \rd x_i \rd z_i \prod_r^{k-1} \rd y^{(r)}_i \delta(\sum_i x_i)\delta(\sum_i z_i)\delta(\sum_i y^{r}_i)\\ \nonumber
\frac{1}{ch(x_i-m_{\rho(i)}) ch(y^{(1)}_i-x_i-M_1 )ch(y^{(2)}_i-y^{(1)}_i-M_2 )
\cdots ch(z_i-y^{(k-1)}_i-M_k)  ch(z_i-\tilde m_i)}.\\
\een

Now we shift $x_i\to x_i + m_{\rho(i)}$, $z_i\to z_i + \tilde m_i$, $y^{(r)}\to \sum^r_j y^{(j)}_i+ x_i + m_{\rho(i)}+\sum^r_j M_j$ and rewrite the partition function as:

\ben 
\label{lq}
\nonumber \Z^{SU(N)^{k+1}}_{2N}&=&\frac{1}{\prod^N_{i<j}sh(m_i-m_j)sh(\tm_i-\tm_j)}\sum_{\rho \in S^N}(-)^{\rho}
\int \prod_i^{N-1} \rd x_i \rd z_i \prod_r^{k-1} \rd y^{(r)}_i \\ \nonumber&&\times
\frac{1}{\prod_i^{N-1} ch(x_i)ch(z_i)  ch(\sum_i^{N-1} x_i+\sum_i^N m_i ) ch(\sum_i^{N-1} z_i+\sum^N_i \tilde m_i )  }
\\&& \times \nonumber\frac{1}{\prod_r^{k-1}\prod_i^{N-1} ch(y^{(r)}_i) ch(\sum_i^{N-1} y^{(r)}_i+ N M_r )
}
\\ &&\times \frac{1}{\prod_i^{N-1} ch(z_i+\tilde m_i-\sum^k_j y^{(j)}_i -x_i-m_{\rho(i)} -\sum_r^k M_r) 
}
\\ &&\!\!\!\!\!\!\!\!\!\!\!\!\!\!\!\!\times \nonumber \frac{1}{ ch(-\sum_i^{N-1}(z_i+\tilde m_i)+\sum_i^{N-1}\sum^k_j y^{(j)}_i +\sum^{N-1}_i(x_i+m_{\rho(i)}) +(N-1)\sum_r^{k-1} M_r-M_k) 
}.
\een

Let's now look at the mirror star-shaped quiver.

The partition function reads:
\ben
\Z^{\times \cdots\times \odot \odot}_N&=&\int \frac{\rd^Nx \delta(\sum x)\prod_{i<j}sh^2(x_i-x_j)}{N!}   {\Z}^{\odot}_N(x; e_i){\Z}^{\odot}_N(x; \tilde{e_i}) \nonumber \\
&& \times \int
  \rd b \rd c \prod_j^k \rd a_j \frac{e^{2\pi i (\eta_b b+\eta_c c)}}{\prod_i^N ch(x_i-b)ch(x_i-c)} \frac{ e^{2\pi i \sum_j^k \xi_j a_j}}{\prod_i^N \prod_j^k ch(x_i-a_j)}. 
\een
Plugging in the exact expression for  $\Z^{T(SU(N)}$  we find:
\ben
&&\frac{1}{i^{N(N-1)} \prod_{i<j}^N sh(e_i-e_j) \prod_{i<j}^N sh(\te_i-\te_j)}\sum_{\rho} (-1)^{\rho}  \int d^N x \delta(\sum x)e^{2 \pi i \sum_j x_j (e_{\rho(j)}+\te_j  )}\nonumber\\
&& \times \int
  \rd b \rd c \prod_j^k \rd a_j \frac{e^{2\pi i (\eta_b b+\eta_c c)}}{\prod_i^N ch(x_i-b)ch(x_i-c)} \frac{ e^{2\pi i \sum_j^k \xi_j a_j}}{\prod_i^N \prod_j^k ch(x_i-a_j)} , 
\een
where we performed a  change of  integration  variables to remove one sum over permutations.

We need  the following integral:
\ben
&&\int  \rd^Nx \delta(\sum x) \rd b \rd c \prod_j^k \rd a_j e^{2\pi i \sum^N_i x_i A_i}\frac{e^{2\pi i (\eta_b b+\eta_c c)}}{\prod_i^N ch(x_i-b)ch(x_i-c)} \frac{ e^{2\pi i \sum_j^k \xi_j a_j}}{\prod_i^N \prod_j^k ch(x_i-a_j)}  =\nonumber \\ &&=
\int  \rd^Nx \delta(\sum x) \rd b \rd c \prod_j^k \rd a_j e^{2\pi i \sum^N_i x_i A_i}e^{2\pi i (\eta_b b+\eta_c c)}e^{2\pi i \sum_j^k \xi_j a_j}\nonumber\\ &&\times
\prod_i^N \int \rd s^{(1)}_i \rd s^{(2)}_i    \frac{ e^{2\pi i (s^{(1)}_i  (x_i-b)  +s^{(2)}_i  (x_i-c)     ) }}{chs^{(1)}_i chs^{(2)}_i }
\prod_i^N  \prod_j^k  \int \rd t^{(j)}_i   \frac{  e^{2\pi i t^{(j)}_i  (x_i-a_j)  }}{ch t^{(j)}_i },
\een
with $ A_i=e_{\rho(i)}+\tilde e_i$.
The integration over $b,c,a_j,x_i$ produces the following deltas:

\ben
&&\delta(-\sum_i^N s^{(1)}_i +\eta_b),\qquad \delta(-\sum_i^N s^{(2)}_i +\eta_c),\\
&&\delta(-\sum_i^N t^{(j)}_i +\xi_j), \qquad j=1,\cdots k,\\
&&\delta(A_i-A_N+s^{(1)}_i-s^{(1)}_N  +s^{(2)}_i-s^{(2)}_N +\sum_j^k t^{(j)}_i  -\sum_j^k t^{(j)}_N  ),\qquad i=1,\cdots N-1
\een

we choose as independent variables $s^{(1)}_i,s^{(2)}_i  $ and $t^{(j)}_i$, $j=1,\cdots k-1$
with $i=1,\cdots N-1$.

From the first three deltas we obtain:

\ben
\nonumber s^{(1)}_N=-\sum_i^{N-1}s^{(1)}_i+\eta_b, \qquad s^{(2)}_N=-\sum_i^{N-1}s^{(2)}_i+\eta_c,\qquad 
 t^{(j)}_N=-\sum_i^{N-1} t^{(j)}_i +\xi_j, \qquad j=1,\cdots k.\\
\een

By manipulating the system we obtain the following equation:

\ben
&&NA_i-\sum^N_l A_l+N s^{(1)}_i-\eta_b + N s^{(2)}_i-\eta_c+
\sum_j^{k-1} ( N t^{(j)}_i-\xi_j)
+ N t^{(k)}_i-\xi_k=0
\een
from which we get:
\ben
-t^{(k)}_i= \frac{- \eta_b-\eta_c -\sum_j^k \xi_j + N A_i -\sum^N_l A_l }{N}+ s^{(1)}_i+ s^{(2)}_i+ \sum_j^{k-1}  t^{(j)}_i, \qquad i=1,\cdots N-1,\nonumber\\
\een

and

\ben
t^{(k)}_N&=&-\sum^{N-1}t^{(k)}_i+\xi_k= \frac{(N-1)( -\eta_b-\eta_c -\sum^N_l A_l ) -(N-1)\sum_j^k \xi_j+\xi_k + N \sum_i^{N-1}A_i}{N}+\nonumber \\ &&+  \sum_i^{N-1}(s^{(1)}_i+ s^{(2)}_i+ \sum_j^{k-1}  t^{(j)}_i).
\een

In terms of these variables the partition function can be rewritten as:

\ben 
\label{kcodine}
\nonumber
\Z^{\times \cdots\times \odot \odot}_N&=&
\frac{1}{i^{N(N-1)}}\frac{\mathcal{J}}{\prod^N_{i<j}sh(e_i-e_j)sh(\te_i-\te_j)}\sum_{\rho \in S^N}(-)^{\rho}
\int \prod_i^{N-1} \rd s^{(1)}_i \rd s^{(2)}_i \prod_r^{k-1} \rd y^{(r)}_i \\ \nonumber&&\times
\frac{1}{\prod_i^{N-1} ch(s^{(1)}_i)ch(s^{(2)}_i)  ch(\sum_i^{N-1} s^{(1)}_i-\eta_b ) ch(\sum_i^{N-1} s^{(2)}_i-\eta_c )  }
\\ \nonumber &&\times  \frac{1}{\prod_j^{k-1}\prod_i^{N-1} ch(t^{(j)}_i) ch(\sum_i^{N-1} t^{(j)}_i- \xi_j )
}
\\ \nonumber && \times \frac{1}{\prod_i^{N-1} ch\left( \frac{ -\eta_b-\eta_c -\sum_j^k \xi_j + N A_i -\sum^N_l A_l }{N}+ s^{(1)}_i+ s^{(2)}_i+ \sum_j^{k-1}  t^{(j)}_i\right)} \\
\nonumber &&\!\!\!\!\!\!\!\!\! \times\frac{1}{ch\left(\frac{(N-1)(- \eta_b-\eta_c -\sum^N_l A_l ) +(N-1)\sum_j^k \xi_j-\xi_k + N \sum_i^{N-1}A_i}{N}+  \sum_i^{N-1}(s^{(1)}_i+ s^{(2)}_i+ \sum_j^{k-1}  t^{(j)}_i)\right)}.\\
\een
Where $\mathcal{J}$ is a constant coming from the Jacobian.
It is easy to see that this expression coincides  with the partition function of the  mirror theory eq. (\ref{lq}), when inserting  the  following dictionary:
\ben
\label{dictio}
\nonumber \eta_b=\sum_i^N m_i,\qquad 
\eta_c= -\sum_i^N \tm,&&\qquad
e_i = -m_i, \quad 
\tilde{e_i}=\tm_i,\\
\xi_j= N M_j,&& \qquad j=1,\cdots k,
\een
notice that the two small tails with FI's $\eta_{b,c}$ carry the extra $U(1)$ flavour symmetry  (with charge $\pm 1$)  of the two full puncture.
While the other $k$ small tails carry the $U(1)$ symmetry  associated to the bi-fundamentals.

\section{$T_N$ theories}\label{nonlagra}
We now move to the study of generalised quiver theories.
The  natural building blocks to construct these theories  are 
partition functions associated to spheres with 3  punctures.
Our first goal will be the computation of the  partition function of the $T_N$ theory $\Z_{T_N}(m_i,\tilde m_j,\hat m_k)$,
associated to the sphere with 3 full punctures,  which depends on three sets of $SU(N)$ masses.
Another block that we will need is the partition function of the {\it ugly} theory 
associated to the sphere with two full and one minimal puncture $\Z(m_i,\tilde m_j, \eta)$, which we 
computed in  section \ref{sfree}.
Finally we  need the block  $\Z(m_i, \eta_a,\eta_b)$ for  the {\it bad} theory
associated to the sphere with one maximal and two minimal punctures
which is given by:
\ben 
\nonumber \Z(m_i, \eta_a,\eta_b)&=&
\frac{1}{i^{N(N-1)/2}\prod^N_{i<j}sh(m_i-m_j)}
\int  \rd^Nz_i \delta(\sum z_i)  \prod_{i<j} sh(z_i-z_j) e^{2 \pi i \sum_j z_j m_j} \nonumber\\
&& \times \int \rd a \rd b \frac{ e^{2 \pi i  (\eta_a a+\eta_b b)}   }{\prod_i^N ch(z_i-a)ch(z_i-b) }.
\een
In section \ref{sasso} we will show how to obtain generalised quiver theories by gluing 
the $\Z(m_i, \eta_a,\eta_b)$, $\Z(m_i,\tilde m_j, \eta)$ and $\Z_{T_N}(m_i,\tilde m_j,\hat m_k)$ blocks.

\subsection{The $T_N$ block}
\begin{figure}[!ht]
\leavevmode
\begin{center}
\includegraphics[height=2cm]{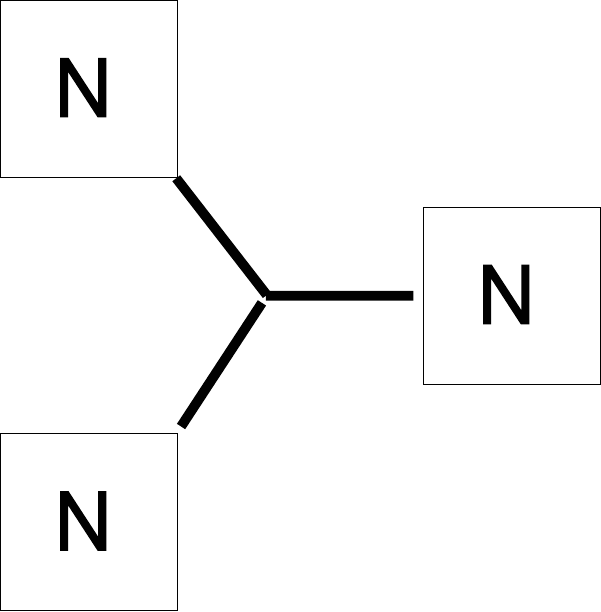} \qquad \includegraphics[height=3cm]{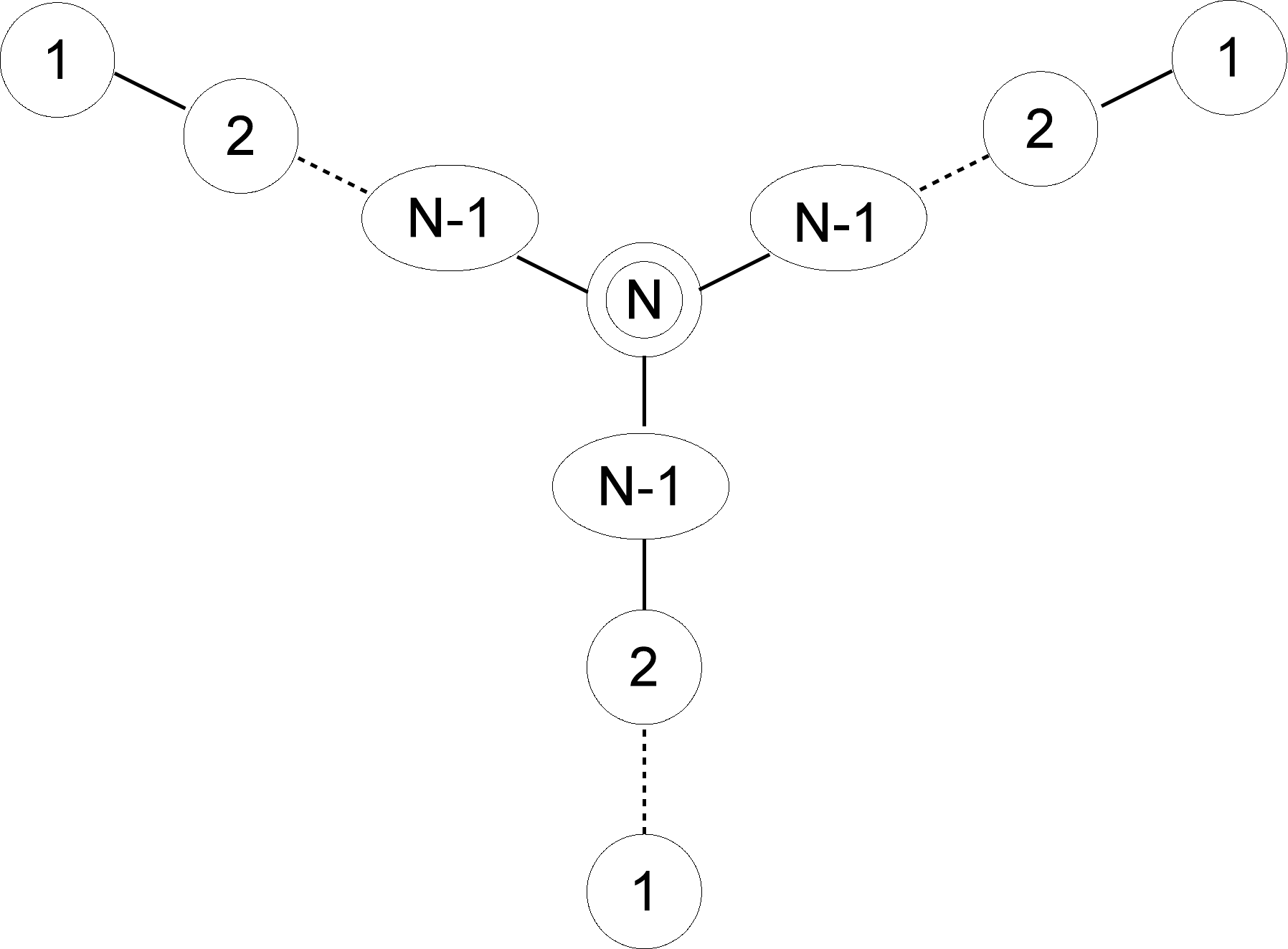}
\end{center}
\caption{The  $T_N$ theory and its mirror.}
\label{tnmtn}
\end{figure}

The $T_N$ theory  is not Lagrangian and in principle one can not use localisation of the path integral to evaluate the partition function,
however, assuming  mirror symmetry we can obtain the $T_N$ partition function from its mirror:
the Lagrangian star shaped quiver theory obtained gluing three $Z^{\odot}_N$ blocks depicted in Fig. \ref{tnmtn}.

The partition function reads:
\ben
\label{imtn}
\nonumber \Z^{\odot \odot \odot}_N(e_i, \tilde e_j, \hat e_k) &=& \int \frac{\rd^Nx \delta(\sum x)\prod_{i<j}sh^2(x_i - x_j)}{N!}\Z^{\odot}_N(x;e_i)\Z^{\odot}_N(x; \tilde e_i)\Z^{\odot}_N(x;  \hat e_k)=\\&&\!\!\!\!\!\!\!\!\!\!\!\!\!\!\!=\nonumber
\frac{1}{i^{3 N(N-1)/2} \prod^N_{i<j}sh(e_i- e_j)sh(\tilde e_i-\tilde e_j)sh( \hat e_i-\hat e_j)}
 \sum_{\rho,\tilde \rho,\hat \rho} (-1)^{\rho+\tilde \rho+\hat\rho} \\ \nonumber
 &&\times \int \frac{d^Nx \delta(\sum x)}{N!\prod_{i<j}sh(x_i - x_j)}
 e^{2 \pi i \sum  x_i (e_{\rho(i)}+\tilde e_{\tilde \rho(i)}+\hat e_{\hat \rho(i)}   ) }.\nonumber\\
 \een
 
 Inserting the dictionary $e_i \to m_i$, $\tilde e_i \to \tilde m_i$ and $\hat e_i \to \hat m_i$ we obtain  the partition function of the $T_N$ theory:

 \ben
\label{itn}
\nonumber \hat \Z_{T_N}(m_i, \tilde m_j, \hat m_k) &= &
\frac{1}{i^{3 N(N-1)/2} \prod^N_{i<j}sh(m_i- m_j)sh(\tilde m_i-\tilde m_j)sh( \hat m_i-\hat m_j)}\\
&&\times \sum_{\rho,\tilde \rho,\hat \rho} (-1)^{\rho+\tilde \rho+\hat\rho}\int \frac{\rd^Nx \delta(\sum x)}{N!\prod_{i<j}sh(x_i - x_j)}e^{2 \pi i \sum  x_i (m_{\rho(i)}+\tilde m_{\tilde\rho(i)}+\hat m_{\hat \rho(i)}   ) }.\nonumber\\
 \een
 
This expression has a  manifest $(S_N)^3$ symmetry.
In particular, thanks to the symmetrisation,  the partition function is finite.
We will see how this  works in detail for the $T_3$ case.

 \paragraph{The $T_3$ theory}

Let's now focus on the  $T_3$ case.
We need to compute  the following divergent integral:
\ben
F(C_i)= \int \frac{\rd^3x \delta(\sum x)}{\prod_{i<j}sh(x_i - x_j)}e^{2 \pi i \sum C_i x_i}
\een
with $C=(C_1, C_2, C_3)=(\{e_{\rho(i)}+\tilde e_{\tilde \rho(i)}+\hat e_{\hat \rho(i)} \})$.
We shift $x_1\to x_1+x_3$ and $x_2\to x_2+x_3$ and get:
\ben
&&  \int \frac{\rd x_1 \rd x_2  \rd x_3\delta(x_1+x_2+3 x_3)}{sh(x_1 - x_2)   sh(x_1)  sh(x_2) }e^{2 \pi i (C_1 x_1+C_2 x_2+\sum^3_i C_i x_3)}=
\frac{1}{3}  \int \frac{\rd x_1 \rd x_2   e^{2 \pi i \left(B_1 x_1+B_2 x_2\right)} }{sh(x_1 - x_2)   sh(x_1)  sh(x_2) }\nonumber\\
&&=\frac{i^3}{24}\int \rd a \rd b \rd c \int \rd x_1 \rd x_2 e^{2\pi i (B_1 x_1 +B_2 x_2)} e^{2\pi i (a (x_1-x_2) +b x_1 +c x_2)}
th(a)th(b)th(c)= \nonumber \\
&&=\frac{i^3}{24}\int da th(a)th(a+B_1)th(a-B_2),
\een
with $B_i=\langle C,h_i\rangle$.
Where we used that: 
\ben
\frac{1}{sinh(x)}=\frac{i}{2}\int \rd a e^{-2\pi i a x }th(a).
\een

In order to compute the last integral we introduce a  $FI$ parameter, playing the role of an IR regulator:
\ben
&&\int \rd a th(a)th(a+B_1)th(a-B_2) e^{2 \pi i \xi a}=\\&&=\frac{ i}{sh(\xi)}\left( - \frac{ch{B_1} ch(B_2) } {sh(B_1) sh(B_2)}+\frac{e^{-2\pi i \xi B_1 }ch(B_1) ch(B_1+B_2) }{sh(B_1) sh(B_1+B_2)}+
\frac{e^{2\pi i \xi B_2 } ch(B_2) ch(B_1+B_2) }{sh(B_1+B_2) sh(B_2) } \right)\nonumber,\een
and expand the result for  $\xi\to 0$
\ben
\label{pole}
=\frac{i}{  2\pi\xi}+  \left(B_1cothB_1-B_2 cothB_2\right)coth(B_1+B_2).
\een
Notice that the divergent term cancels out   thanks to the sum over the 3 sets of $S_3$ permutations.
The on-shell\footnote{On-shell $\sum_i m_i=0$ and $\langle m,h_i\rangle=m_i$. } partition function reads:

\ben
\nonumber \Z_{T_3}(m_i, \tilde m_j, \hat m_k)& = &
\frac{1}{3!  \prod^3_{i<j}sh(m_i- m_j)sh(\tilde m_i-\tilde m_j)sh( \hat m_i-\hat m_j)}
 \sum_{\rho,\tilde \rho,\hat \rho}(-1)^{\rho+\tilde \rho+\hat\rho} \\&&\!\!\!\!\!\!\!\!\!\!\!\!\!\!\!\!\!\!\!\!\!\nonumber
\times (m_{\rho(1)}+\tilde m_{\tilde \rho(1)}+\hat m_{\hat \rho(1)}) coth(m_{\rho(1)}+\tilde m_{\tilde \rho(1)}+\hat m_{\hat \rho(1)} )  coth(m_{\rho(3)}+\tilde m_{\tilde \rho(3)}+\hat m_{\hat \rho(3)} ).\\
 \een

\subsection{Consistency checks from S-duality invariance}\label{sasso}
We will now glue our building blocks to obtain generalised quiver theories associated to spheres with arbitrary punctures.
The partition functions we will construct must satisfy an important consistency condition:
they must be independent on the particular pants-decomposition we choose to perform the gluing.
This is a  consequence of the fact that our theories are independent on the complex structure of the punctured  sphere
and they have the structure of a  2d TQFT.
This has been recently pointed out in  \cite{gadde}.
The super-conformal index of a  $4d$ theory on a  punctured Riemann surface,
 which  is computed by a  2d TQFT   \cite{rasta1,rasta2}, has been shown to reduce,
  in a certain limit, to the 3d partition function associated to  the same punctured Riemann surface \cite{spiri,gadde}.
It is then expected that the 3d partition function will inherit the TQFT structure from the index.

\begin{figure}[!ht]
\leavevmode
\begin{center}
\includegraphics[height=4.9cm]{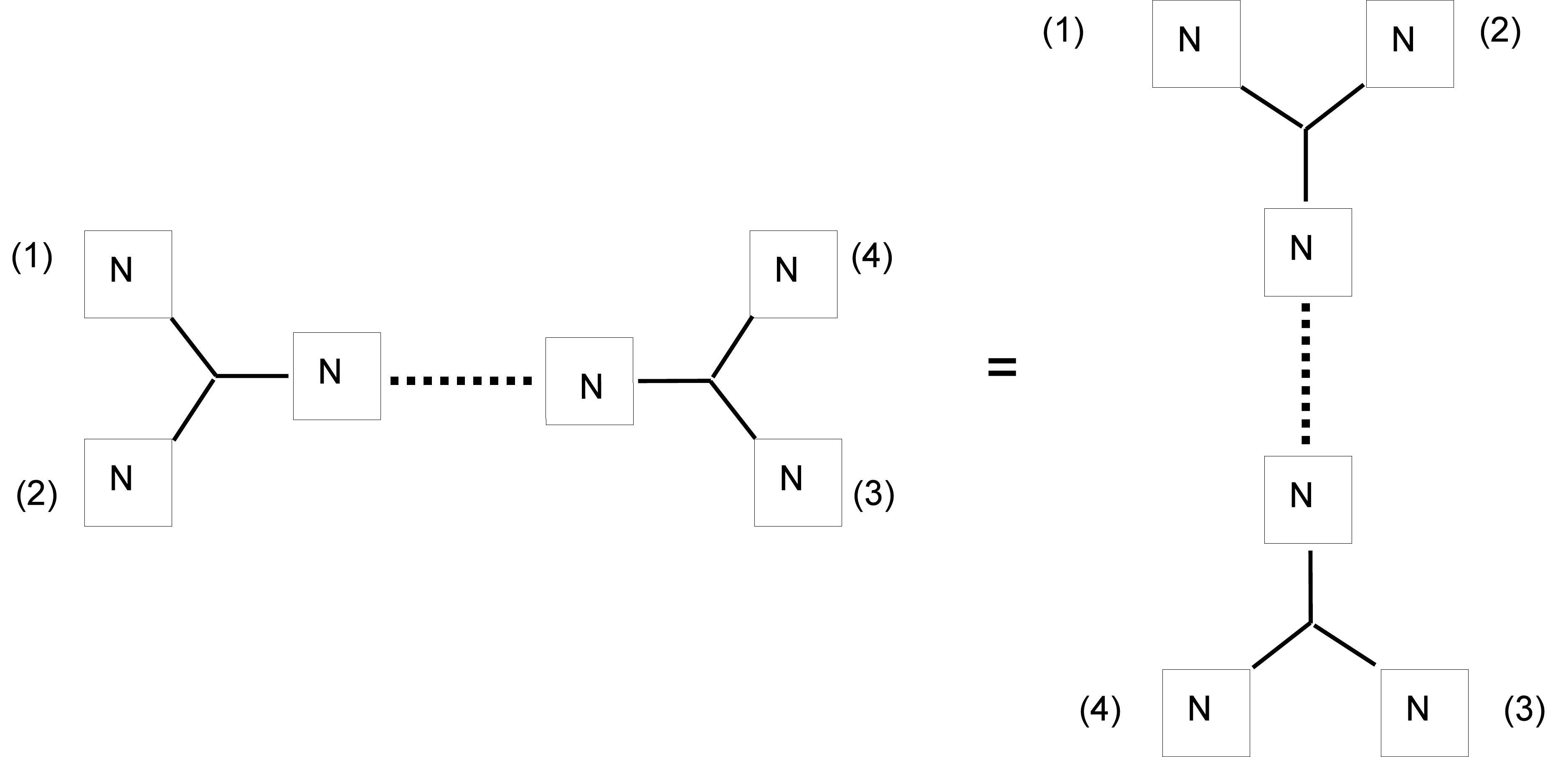}
\end{center}
\caption{S-duality invariance as associativity relation for $T_N$ blocks.}
\label{asso}
\end{figure}

To test our blocks, we will show that they satisfy  the operator algebra of a 2d TQFT,
 in particular we will prove the associativity relation indicated in Fig. \ref{asso},
stating that the partition function of the sphere with four full punctures can be obtained equivalently as:
 \ben
\label{associativity}
\nonumber \!\!\!\! \Z(m^{(1)}_i,  m^{(2)}_j, m^{(3)}_k, m^{(4)}_l)&=&
`` \sum_{y_n}"\Z_{T_N}(m^{(1)}_i,  m^{(2)}_j, y_n)\Z_{T_N}(y_n,m^{(3)}_k,  m^{(4)}_l)=\\&=& ``\sum_{y_n}"\Z_{T_N}(m^{(1)}_i,  m^{(4)}_j, y_n)\Z_{T_N}(y_n,m^{(3)}_k,  m^{(2)}_l)
 =\Z(m^{(1)}_i,  m^{(4)}_j, m^{(3)}_k,m^{(2)}_l).\nonumber\\
 \een
 

To glue two $T_N$ blocks we simply gauge one of the $SU(N)$ flavour symmetry and we get:
 \ben
&&\nonumber\!\!\!\!\!\!\!  \!\!\!\! \!\!\!\! \Z(m^{(1)}_i,  m^{(2)}_j, m^{(3)}_k, m^{(4)}_l) = \int \frac{\rd^Ny_i \delta(\sum y_i)\prod_{i<j}sh^2(y_i - y_j)}{N!}
\Z_{T_N}(m^{(1)}_i,  m^{(2)}_j, y_n)\Z_{T_N}(y_n,m^{(3)}_k,  m^{(4)}_l)=
\\  &&=\nonumber
\frac{ \sum_{\rho^{(1)},\rho^{(2)} , \rho^{(3)} ,\rho^{(4)}}   (-1)^{\rho^{(1)}+\rho^{(2)} + \rho^{(3)} +\rho^{(4)}  }}{  \prod^N_{i<j}sh(m^{(1)}_i- m^{(1)}_j)  sh(m^{(2)}_i- m^{(2)}_j)sh(m^{(3)}_i- m^{(3)}_j)sh(m^{(4)}_i- m^{(4)}_j) }\\
 \nonumber&&\times
\nonumber \int \frac{\rd^Ny_i \delta(\sum y_i)}{N!}\!\!\!
 \int \frac{\rd^Nx \delta(\sum x)}{\prod_{i<j}sh(x_i - x_j)}e^{2 \pi i \sum  x_i (m^{(1)}_{\rho^{(1)}(i)}+ m^{(2)}_{\rho^{(2)}(i)}+y_i   ) }\\ \nonumber &&\times
  \int \frac{\rd^Nz \delta(\sum z)}{\prod_{i<j}sh(z_i - z_j)}e^{2 \pi i \sum  z_i (m^{(3)}_{\rho^{(3)}(i)}+m^{(4)}_{\rho^{(4)}(i)}+y_i   ) }.\nonumber\\
 \een
 The integration over $y_i$ produces a delta function setting  $z_i=x_i$ and we obtain:
 \ben
&&\frac{1}{  \prod^N_{i<j}sh(m^{(1)}_i- m^{(1)}_j)  sh(m^{(2)}_i- m^{(2)}_j)sh(m^{(3)}_i- m^{(3)}_j)sh(m^{(4)}_i- m^{(4)}_j) }\nonumber\\&&
 \sum_{\rho^{(1)},\rho^{(2)} , \rho^{(3)} ,\rho^{(4)}}  (-1)^{  \rho^{(1)}+\rho^{(2)} + \rho^{(3)} +\rho^{(4)}  }\times \int \frac{\rd^Nx_i \delta(\sum x_i)}{N!}
  \frac{e^{2 \pi i \sum  x_i (m^{(1)}_{\rho^{(1)}(i)}+ m^{(2)}_{\rho^{(2)}(i)}+m^{(3)}_{\rho^{(3)}(i)}+m^{(4)}_{\rho^{(4)}(i)}     )}}{\prod_{i<j}sh(x_i - x_j)^2 }.\nonumber\\
 \een
This expression is manifestly invariant under permutations of the $m^{(I)}$'s  and thus the associativity 
property eq. (\ref{associativity})  is satisfied.

\begin{figure}[!ht]
\leavevmode
\begin{center}
\includegraphics[height=2cm]{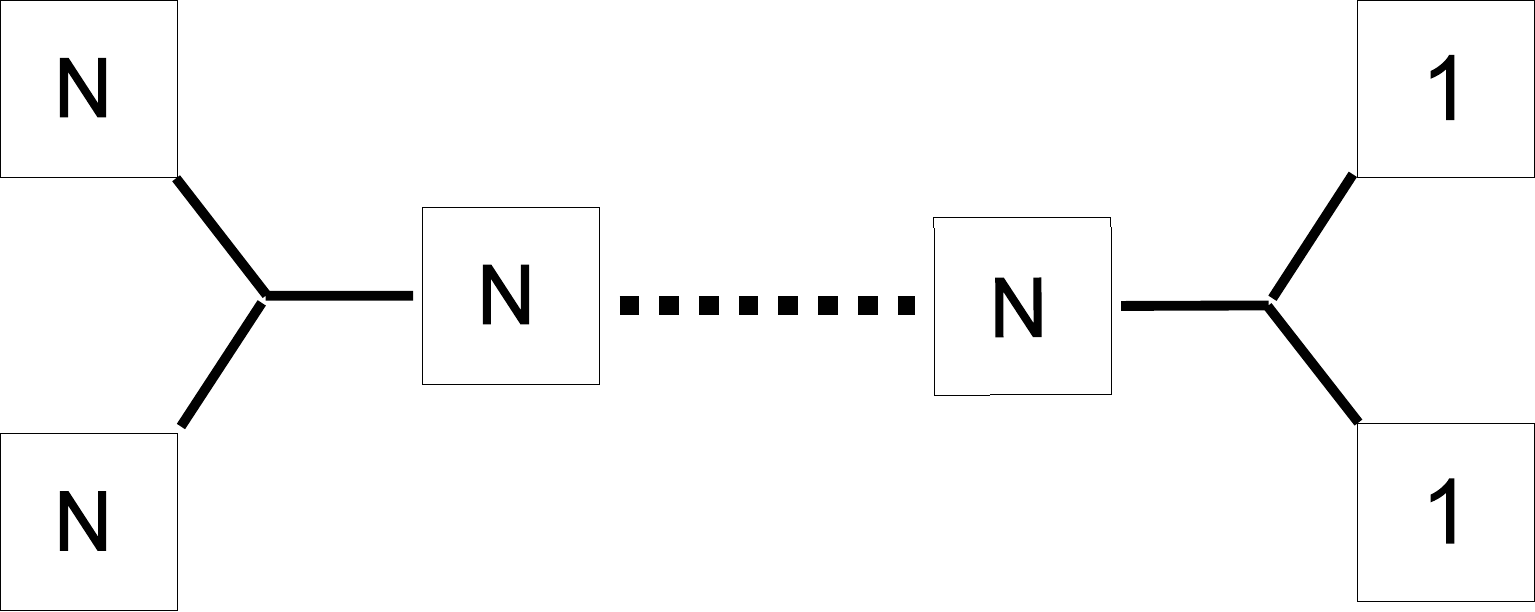}\qquad \includegraphics[height=2cm]{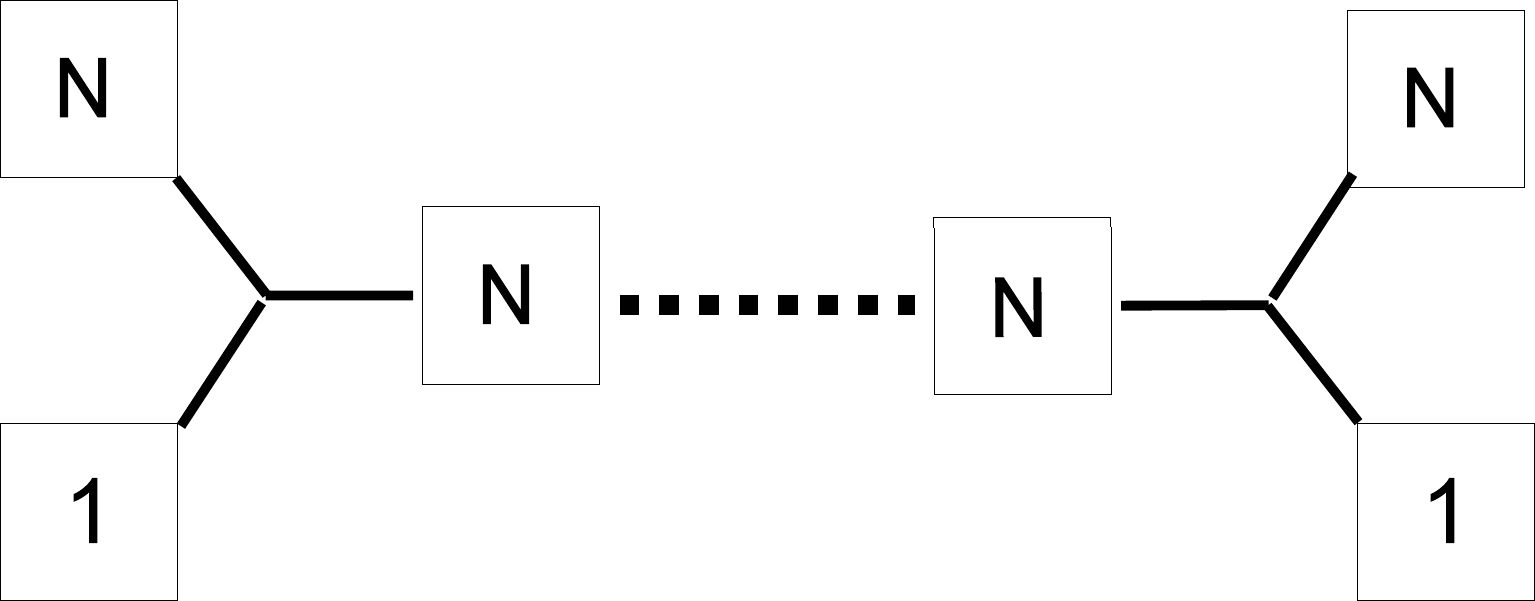}
\end{center}
\caption{Two equivalent  gluing for the $SU(N)$  theory with $N_f=2N$,}
\label{2ways}
\end{figure}

As a further test we show  that  $SU(N)$ theory with $N_f=2N$ can be obtained in two ways.
The first way, depicted on the left  in Fig. \ref{2ways}, corresponds to
 gluing a $T_N$ block $\Z_{T_N}(m_i, \tilde m_j, y_k)$ and a {\it bad} block $ \Z(y_k,\eta_a,\eta_b)$ with
$\eta_a=\sum_i m_i$, $\eta_b=\sum_i \tilde m_i$.
The second way,  depicted on the right  in Fig. \ref{2ways}, corresponds to
 gluing  two {\it ugly} blocks $\Z(m_i,  \eta_a,y_k)$ and $ \Z(y_k,\eta_b,\tilde m_j)$.

Let's start with the first gluing:
\ben
&&\nonumber \Z^{SU(N)}_{2N}(m_i,\tilde m_j)= \int \frac{\rd^Ny_i \delta(\sum y_i)\prod_{i<j}sh^2(y_i - y_j)}{N!}
\Z_{T_N}(m_i, \tilde m_j, y_k)\nonumber \Z(y_k,\eta_a,\eta_b)
=\\ &&=
\frac{1}{ i^{ N(N-1)}   \prod^N_{i<j}sh(m_i- m_j)sh(\tilde m_i-\tilde m_j)}
 \sum_{\rho,\rho'} (-1)^{\rho+\rho'}
 \int \frac{\rd^Ny_i \delta(\sum y_i)}{N!}\nonumber\\
 &&\!\!\!\!\!\!\!\!\! \times\nonumber
 \int \frac{\rd^Nx \delta(\sum x)}{\prod_{i<j}sh(x_i - x_j)}e^{2 \pi i \sum  x_i (m_{\rho(i)}+\tilde m_{\rho'(i)}+y_i   ) }\int \frac{\rd a \rd b \rd^N z_i \delta(\sum z_i) e^{2 \pi i  (\eta_a a+\eta_b b)} e^{2 \pi i \sum_j z_j y_j}  \prod_{i<j} sh(z_i-z_j)}{\prod_i^N ch(z_i-a)ch(z_i-b) },\\
 \een
 the integration  over $y_i$ sets $x_i=z_i$ and we obtain:
 \ben
 \label{test}
\frac{1}{ i^{ N(N-1)}   \prod^N_{i<j}sh(m_i- m_j)sh(\tilde m_i-\tilde m_j)}
 \sum_{\rho,\rho'} (-1)^{\rho+\rho'}
 \int \frac{\rd^Nx_i \delta(\sum x_i)}{N!}\frac{
e^{2 \pi i \sum  x_i (e_{\rho(i)}+\tilde e_{\rho'(i)} )}
 e^{2 \pi i  (\eta_a a+\eta_b b)} }{\prod_i^N ch(z_i-a)ch(z_i-b) }.\nonumber\\
 \een
  
For the second gluing we get:
  
 \ben
&&\nonumber \Z^{SU(N)}_{2N}(m_i,\tilde m_j) = \int \frac{\rd^Ny_i \delta(\sum y_i)\prod_{i<j}sh^2(y_i - y_j)}{N!}
\Z(m_i,\eta_a,  y_k)\nonumber \Z(y_k,\eta_b,\tilde m_j)
=\\ &=&
\frac{1}{ i^{ N(N-1)}   \prod^N_{i<j}sh(m_i- m_j)sh(\tilde m_i-\tilde m_j)}
 \sum_{\rho,\rho'} (-1)^{\rho+\rho'}
 \int \frac{\rd^Ny_i \delta(\sum y_i)}{N!} \nonumber\\
 &&
\times \nonumber  \int \rd^Nx \delta(\sum x)         e^{2 \pi i \sum  x_i (m_{\rho(i)}+y_i   ) } \frac{e^{2\pi i \eta_a a}}{ch(x_i-a)}  \int \rd^Nz  \delta(\sum z)      
    e^{2 \pi i \sum  z_i (\tilde m_{\rho'(i)}+y_i   ) } \frac{e^{2\pi i \eta_b b}}{ch(z_i-b)},  \\
 \een
 integrating over $y_i$ we obtain a delta leading to $x_i=z_i$ leading again to the result in eq. (\ref{test}).

\section{Conclusions}

In this paper we developed a complete formalism to compute partition functions of generalised three-dimensional  quiver theories
deformed by mass and FI parameters. 
We used the mirror description in terms of Lagrangian star shaped quivers combined with  localisation techniques.

One of our main results is  the explicit evaluation of the partition function of the $T(SU(N))$ quiver  theory 
as  a function of the  FI and mass parameters. The  $T(SU(N))$ tail, mirror of the full puncture,
is the fundamental building block to evaluate the partition function of generic star shaped quiver theories.

We provided several non-perturbative checks  of the mirror realisation  in terms of star shaped quivers \cite{sici} by showing that  partition functions of mirror pairs of Lagrangian theories, are equal provided we exchange masses and FI's.

We then assumed mirror symmetry to find the partition function of non-Lagrangian theories in terms of the star-shaped mirrors.
In particular we computed the partition function of the $T_N$ theory  giving an explicit result for the $T_3$ case.

In this paper we only consider full punctures or minimal punctures.
It is however very simple to extend  our results to the case where 
 punctures specified by  generic Young tableaux with N boxes.

An interesting extension of our work would be to evaluate expectations values of supersymmetric observables such as Wilson Loops in the 
3d generalised quiver  theories. With our explicit results for partition functions
it should be possible to determine the mirror dual of these observables.

\section*{Acknowledgments}
S.P would like to thank G.~Bonelli, A.~Brini and F.~Passerini  for useful comments on the draft.
The work of S.P. is supported by a 
Marie Curie Intra-European Fellowship: FP7-PEOPLE-2009-IEF.

  \appendix

\end{document}